\documentclass[aps,prb,twocolumn,superscriptaddress]{revtex4-1}

\usepackage{graphicx}
\usepackage{hyperref}
\usepackage{url}
\usepackage{amsmath}
\usepackage{amssymb}
\usepackage{verbatim}
\usepackage{xcolor}
\usepackage{enumerate}
\usepackage[normalem]{ulem}

\def\beq {\begin{equation}}
	\def\eeq {\end{equation}}

\def\bfk {\mathbf{k}}

\newcommand{\refeq}[1]{Eq.~\eqref{#1}}

\newcommand{\refsec}[1]{Sec.~\ref{#1}}

\newcommand{\br}{{\bf r}}

\begin{document}

\title{Re-using model results to determine materials properties: connector theory approach}

\newcommand{\lsi}{LSI, CNRS, CEA/DRF/IRAMIS, \'Ecole Polytechnique, Institut Polytechnique de Paris, F-91120 Palaiseau, France}
\newcommand{\etsf}{European Theoretical Spectroscopy Facility (ETSF)}
\newcommand{\soleil}{Synchrotron SOLEIL, L'Orme des Merisiers, Saint-Aubin, BP 48, F-91192 Gif-sur-Yvette, France}
\newcommand{\jku}{Johannes Kepler University Linz, Austria}
\newcommand{\epfl}{Theory and Simulation of Materials (THEOS), \'Ecole Polytechnique F\'ed\'erale de Lausanne, CH-1015 Lausanne, Switzerland}

\author{Marco Vanzini}
\affiliation{\lsi}
\affiliation{\epfl}
\affiliation{\etsf}
\altaffiliation{M.V. and A.A. contributed equally to this work.}
\email{marco.vanzini@epfl.ch}

\author{Ayoub Aouina}
\affiliation{\lsi}
\affiliation{\etsf}
\altaffiliation{M.V. and A.A. contributed equally to this work.}

\author{Martin Panholzer}
\affiliation{\lsi}
\affiliation{\etsf}
\affiliation{\jku}

\author{Matteo Gatti}
\affiliation{\lsi}
\affiliation{\etsf}
\affiliation{\soleil}

\author{Lucia Reining}
\affiliation{\lsi}
\affiliation{\etsf}
\email{lucia.reining@polytechnique.fr}

	\begin{abstract}
		
		Computational materials design often profits from the fact that some
		complicated contributions are not calculated for the real material, but replaced by results of models. We turn this approximation into a very general and in principle exact theory by introducing the concept of a connector, which is a prescription of how to use the results of a model system in order to simulate a real system. We set the conditions that must be fulfilled for the existence of an exact connector. 
		We demonstrate that, and why, this approach is a very convenient starting point for approximations.
		We also show that the connector theory can be used to design new functionals, for example for density functional theory.  
		We illustrate our purposes with simple but realistic examples.
		%{\color{blue}Don't forget: s a potential described by two parameters can still be seen as a series of values in space points that are then averaged or so. Also: when is direct better than connector? For example, if approx is exact in real but not in model, maybe because higher orders disappear in fluctuations (work it out)}
		
	\end{abstract}
	
	
	\maketitle
	
	\section{Introduction}
	
	Computational materials design \cite{Curtarolo2013,Saal2013,Marzari2016} is complicated by the complexity of materials and by interaction effects. 
	This hampers both calculations and understanding. 
	The fundamental problem lies in the fact that the effects of the Coulomb interaction and of the specific material cannot be separated. Otherwise, 
	one could calculate the interaction contributions once and for all, store them and add them every time a new  material is calculated. This is an unreachable dream, but it still indicates an intriguing direction of thinking: in some model systems the effects of the Coulomb interaction can be treated exactly, or at least to a much better extent than in real systems, and
	attempts to use model results in order to simulate interaction effects in real materials are widespread. The most prominent example is the local density approximation (LDA) to the Kohn-Sham (KS) exchange-correlation (xc) potential $v_{\rm xc}(\br ;[n])$ of density functional theory (DFT) \cite{DFT1}. The exact potential is unknown in most real materials. The LDA replaces $v_{\rm xc}(\br ;[n])$  at a point ${\bf r}$ by its value in a homogeneous electron gas (HEG) that is calculated at the density $n({\bf r})$ of the real system in the same point ${\bf r}$. %\textit{instead of:}}}\textit{The most prominent example is the local density approximation (LDA) to density functional theory (DFT) \cite{DFT1}, where the Kohn-Sham (KS) exchange-correlation (xc) potential $v_{\rm xc}(\br ;[n])$ at a point ${\bf r}$  is taken from a homogeneous electron gas (HEG) that is calculated at the density $n({\bf r})$ of the real system in the same point ${\bf r}$.} 
	In this way, DFT profits from the existence of tabulated and interpolated Quantum Monte Carlo results \cite{Cepe1980}. Similarly, dynamical mean field theory in the single site approximation  takes the effective local self-energy from the Anderson impurity model, and although in this case results have not been tabulated, the procedure has enabled a realistic description of correlated materials \cite{RevDMFT}. However, in spite of numerous studies and attempts \cite{DFT1,Fermi1927,Thomas1927,Ma1968,Vosko1980,Gunnarsson1977,Alonso1977,Alonso1978,Gunnarsson1979,Gunnarsson1980,Cuevas2012,LangrethPerdew1980,PerdewWang1986,PBE1996,Perdew1999,Perdew2008,Sun2015,Dobson-book1998,Jung2004,Gonzalez1996,Olsen2012,Olsen2014,Patrick2015,Schmidt2017,Lu2014,Vignale1987,Zangwill1980,GrossKohn1985,Dobson1997,Ng1989,VignaleKohn1996,VignaleUllrichConti1997,Tokatly2007,Gao2010,Nazarov2010,Trevisanutto2013,Panholzer2018,giuliani2005quantum,martin2004} it is very difficult to go beyond these simple schemes.  One reason is that using results of the models  is considered from the very beginning as an \textit{approximation}.
	%\footnote{A local, but in principle exact framework for DMFT, called ``spectral density functional theory'', has been proposed \cite{Savrasov2004}. However, up to now no useful approximation has been found.}.
	Breakthrough is instead often based
	on an \textit{exactification}: a recent example is the exact factorization
	of the many-body wavefunction \cite{BO_Gross,BO_Gross2,Gidopoulos} that includes the Born-Oppenheimer approximation as limiting case \cite{Eich2016} and allows developments beyond it \cite{Min2015}.
	In a similar spirit, one may try to formulate a general, in principle exact, approach that uses model results for building a density functional for $v_{\rm xc}({\bf r};[n])$, of which the LDA would be one particular approximation.
	
	We keep this idea as broad as possible and therefore pose the following questions:
	\textit{Can one exactify the idea of re-using 
		results from one system, for example a model, to describe another system? If yes, under which conditions? And does this suggest strategies for systematic approximations?} These are the  questions at the heart of our work. Our answer, termed
	\textit{connector theory} (COT), tells how one can in principle connect different systems, and how to find good approximations in practice. It is very general and designed to overcome numerous obstacles, not  only the problem of interaction. It has potential for great speedup of calculations, and, as we will demonstrate, it is a powerful tool to design improved functionals, for example, but not exclusively, for DFT.
	
	\section{Formalism}
	The idea can be summarized as follows: 
	%Formally, the kind of problems which this work is aimed at is the following: 
	suppose one wishes to calculate 
	an observable or quantity $O$ that is a function or functional of a set of parameters or function $Q$, and it can depend on additional arguments $x$, so $O=O(x;Q)$. For example, if $O$ is the xc potential $v_{\rm xc}({\bf r};[n])$, then $x$ is a spatial coordinate ${\bf r}$ and $Q$ is a function, the density $n({\bf r}')$. Or, $O$ could  be a
	spectral function, $x$ the frequency, and $Q$ the atomic potentials, etc. In most cases $O(x;Q)$ is difficult or impossible to calculate in a real material without approximations, or even unknown. However, it may be possible to calculate $O(x;Q)$ for some  $Q$ in a restricted domain:  this restricted domain defines a model. For example, when $Q$ is the electron density and we restrict it to homogeneous densities, the model is the HEG. Connector theory aims at using the model results in
	order to simulate systems where $Q$ lies outside the model domain. 
	%The underlying hypothesis is that not all details of
	%the parameters that describe the real system are equally
	%important. 
	The goal is therefore to find, for a given real system where $Q\equiv Q^R$, another $Q \equiv Q^c$ that lies in the model domain, such that 
	\begin{equation}
		O(x;Q^R)=O(x;Q^c_x)\,.
		\label{eq:connector}
	\end{equation}
	Note the subscript $x$ of $Q^c_x$, which indicates that $Q^c$ is allowed to be different for every value of the argument $x$.
	
	In the following we suppose that, because the model is simple, it is described by only one effective parameter $\mathcal Q$  or, when there are more parameters, we can choose one that will be used to fulfill  (\ref{eq:connector}). This restriction can be dropped, but it is often useful and we keep it here for clarity. In this case, in the model the functional or multi-dimensional function can be represented by a simple function,  $O(x;Q^c_x)\to \mathcal {O}_x(\mathcal Q^c_x)$. For example, in the HEG the one parameter $\mathcal Q$ is its number density $ n^h$, so  $v_{\rm xc}({\bf r};[n])\to v_{\rm xc}^h(n^h)$, where $v_{\rm xc}^h$ corresponds to $\mathcal {O}$ and $n^h$ to $\mathcal Q$. Eq. \eqref{eq:connector} reads then
	\begin{equation}
		O(x;Q^R)=\mathcal {O}_x(\mathcal Q^c_x)\,.
		\label{eq:connector-mathcal}
	\end{equation}
	
	Of course, even allowing $\mathcal Q^c_x$ to depend on $x$ it could be impossible to fulfill equality \eqref{eq:connector-mathcal} if one restricts the model domain too severely. However, if the model is flexible enough such that the equation can be satisfied in principle, one can try to find the one or more $\mathcal Q^c_x$ for which the equality holds. 
	In other words, for the idea to be applicable the following condition must be satisfied:
	\begin{itemize}
		\item {\bf [A]} On its domain of definition, 
		%$O(x;Q^c_x)$ or the function 
		$\mathcal O_x(\mathcal Q^c_x)$ must  yield all values that 
		%the function or functional 
		$O(x;Q^R)$ can take on its domain, i.e., the $Q^R$ of interest.
	\end{itemize}
	The $Q^R$ which define the domain of interest depend on the range of physical systems one wants to explore; this range does not necessarily include \textit{all possible} physical systems. The domain of $ O$, on the other hand, defines the model system. If for certain $Q^R$ and/or $x$ Eq. \eqref{eq:connector-mathcal} cannot be fulfilled, we have to change model  by changing its domain, i.e. the range of allowed $\mathcal Q^c_x$. 
	
	Eq. (\ref{eq:connector-mathcal}) is then formally solved,
	\begin{equation}
		\mathcal Q^c_x=\mathcal{O}_x^{-1}(O(x;Q^R)).
		\label{eq:inverse}
	\end{equation} 
	This operation requires inversion of $\mathcal{O}_x$, which brings us to a second condition. Indeed, 
	${\bf [A]}$ is the only necessary condition,
	but there is also a question of uniqueness in \eqref{eq:inverse}:  $\mathcal O^{-1}$ may require boundary conditions in order to be well defined.
	This is not a problem of principle, but may  create difficulties for the design of approximations.
	%that rely on \eqref{eq:inverse}. 
	We therefore require:
	\begin{itemize}
		\item {\bf [B]} When the inverse $\mathcal O^{-1}$ of $\mathcal O$ is not unique, it should at least be possible to specify a unique choice among the possibles $\mathcal O^{-1}_i$ .
	\end{itemize}
	
	Finally, the resulting $Q^c_x$ is used to get the observable in the real system using the model,
	%{\color{brown} I'd change 'result': looks like (3) is the result of (2), while it's more the opposite, and (3) is our prescription/recipe} is
	\begin{equation}
		O(x;Q^R)= \mathcal{O}_x(\mathcal Q^c_x).
		\label{eq:final}
	\end{equation}
	%In other words, if a model $ \mathcal Q^c_x$ can be found from \eqref{eq:inverse},
	%or which \refeq{eq:final} is fulfilled, 
	%one can use the results of the model to obtain the quantity of interest in the real system. 
	Therefore, we call $\mathcal{Q}^c_x$ the \textit{connector}.
	The scheme suggests to store $\mathcal O_x$ as numerical data, interpolated analytic expression or, in some cases, analytic result. Once a connector $\mathcal Q^c_x$  is given, for any real system one can then simply use these data instead of
	calculating $O(x;Q^R)$, which will make calculations extremely efficient. This concept of re-using data has made the LDA a breakthrough, since numerous calculations could be performed without ever re-evaluating the interaction effects in the HEG.
	%\textcolor{blue}{Fine for me! }\textcolor{red}{OLD: } tabulate $\mathcal O_x$, {\color{blue} or, if possible, even interpolate it as a function of its variables $x$,} and, once 
	%The advantage of such an approach is clear: once the table $\mathcal Q$ is stored and a connector $\mathcal Q^c_x$  is given, for any real system to simply use these data instead of calculating $O(x;Q^R)$, which will make calculations extremely efficient. {\color{blue} This was exactly the breakthrough that allowed LDA to be used by everyone in the community, without passing each time by the evaluation of $v_{xc}$ in the HEG.} \textcolor{red}{END OLD}
	%route is promising for three reasons: approximations are suggested by the structure of the equations, exact constraints can be included, and the interpretation of results relies on a decomposition in physically meaningful building blocks. 
	
	In practice, however, conditions {\bf [A]} and {\bf [B]} settle the framework, but nothing has been gained yet: the unknown $O(x;Q^R)$ still enters the calculation of the connector %$O_x^c=\mathcal O_x^{-1}(O(x;Q^R))$
	in \eqref{eq:inverse}. Therefore, for COT to be of practical interest we have to add:
	\begin{itemize}
		\item {\bf [C]} The COT ansatz  must suggest a strategy for approximations.
	\end{itemize}
	%There could be more than one solution, but this is not a problem in principle. The link between the real material and the model table is made by $\tilde Q^c_x$: we call it the \textbf{connector}. 
	The LDA can be seen as a connector approximation, where the right hand side of \eqref{eq:inverse} is replaced by the local density,
	$\bar n^c_{\bf r}\approx n({\bf r})$, so $v_{\rm xc}({\bf r};[n])
	\approx v_{\rm xc}^h(\bar n^c_{\bf r})$ according to \eqref{eq:final}. It was based on Kohn's ingenious intuition of nearsightedness \cite{DFT1,Kohn1996,Prodan2005}.  
	%=v_{\rm xc}^h(n({\bf r}))$ 
	A major topic of the present work is to lie out a strategy for systematic approximations, and to generalize the framework to applications well beyond DFT.
	
	For this, we have to prove that \eqref{eq:connector} is a clever starting point. The fundamental idea is to make \textit{in the calculation of the connector} \eqref{eq:inverse} an \textit{equivalent} approximation to $O$ and $\mathcal O$,
	\begin{equation}
		\mathcal Q^{c,{\rm approx}}_x=(\mathcal{O}^{\rm approx}_x)^{-1}(O^{\rm approx}(x;Q^R)).
		\label{eq:inverse-approx}
	\end{equation}
	The final approximate connector result is obtained as
	\begin{equation}
		O(x;Q^R)\approx O^c_x\equiv \mathcal O(\mathcal Q_x^{c{,\rm approx}}).
		\label{eq:connector-result}
	\end{equation} 
	Note that the model observable $\mathcal O$ is supposed to be well known, but it is important to approximate it in \eqref{eq:inverse-approx} in the same way as $O$ of the real system, whereas the exact model function $\mathcal O$ is used in   \eqref{eq:connector-result}. In this way, the result becomes exact in two limits: when  the approximation is increasingly good, even for a very restricted model, and when the domain of the model system tends towards that of the real system, even using a very poor approximation.   
	In practice, one will not be able to do the exact calculation, but away from these limits, using the equivalent approximation for $O^{\rm approx}(x;Q^R)$ and $\mathcal O^{\rm approx}_x(\mathcal Q)$ still leads to error cancelling, and the limiting behaviour indicates that results can be improved in a controllable way. 
	How far the model system can be chosen from the real system depends on the quality of the approximation, and vice versa, how rough the approximation is that one can tolerate depends on the closeness of the model and the real system. This double dependence is a source of the power of the connector approach.  It implies that one can expect the approximate connector result \eqref{eq:connector-result}
	to be superior to the direct approximation $O^{\rm approx}(x;Q^R)$, for similar computational cost.
	This benefit will be larger when the model contains important features of the real system, like in the examples below,  the same kinetic energy, or the same Coulomb interaction. Therefore, in COT the main effort and intuition will go into the choice of a suitable model.
	%Several functionals that came after the LDA can also be interpreted as approximations to connector theory, although they were derived on different grounds.
	
	%\begin{figure}
	%\includegraphics[width=\columnwidth]{scheme3.pdf}
	%\includegraphics[width=\columnwidth]{scheme_4.pdf}
	%   \includegraphics[width=\columnwidth]{scheme_5.pdf}
	%    \caption{Sketch of the potential in Eq. \eqref{eq:pot} for $L=1$ and different values of $\alpha$. In red, the lowest energy levels. \textcolor{red}{These are chosen in such a way that I'm sure the energies are correct, or, we can put the one on the right, that spans all the range in the plot but only for the same energy levels, and one can compare...}}
	%  \label{fig:scheme}
	%\end{figure}

	\section{A simple illustration: energy levels in a 1D potential}
	\label{sec:1Dpot}
	
\begin{figure}  
		\includegraphics[width=\columnwidth]{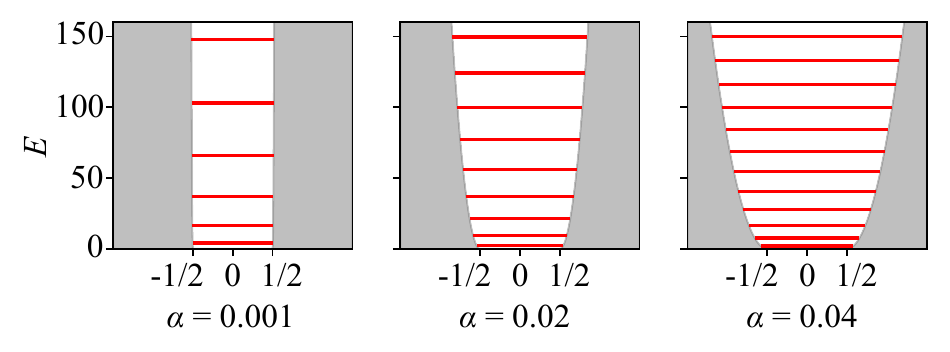}
		\includegraphics[width=\columnwidth]{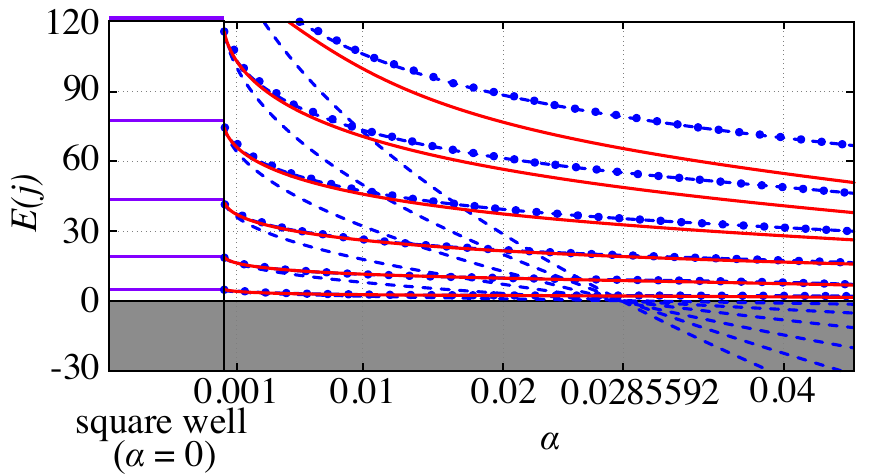}
		\caption{Energy levels in a one-dimensional potential. Top panel: in grey, sketch of the potential $V(z)$ from Eq. \eqref{eq:pot}, as a function of $z$ for $L=1$, and different values of $\alpha=1/L^2\omega_0$. In red, the lowest energy levels $E(j)$. The first six of these are shown in the bottom panel, as a function of $\alpha$: the continuous red line is the exact result. Blue dashed, $E^{\rm approx}_j$ of first-order perturbation theory. 
			%    of the equation of motion. 
			Blue dashed line with symbols, connector results $\mathcal E_j(L^{c,approx}_j)$. Left panel, the  levels $\mathcal E_j(L)$ of an infinite potential well of width $L=1$, 
			%$L^2E(j)=\frac{\pi^2j^2}{2}\frac{\hbar^2}{m}$, 
			which is the $\alpha\to 0$ result.}
		\label{fig:negative}
	\end{figure}

	Let us start with a very simple case to illustrate both the generality of the idea well outside DFT-LDA,  and the improvement obtained by using a connector. Take
	one electron in a one-dimensional potential of shape 	
	\begin{equation}
		V(z)=
		\begin{cases}
			\frac{1}{2}\omega_0^2\left(z+\frac{L}{2}\right)^2 & z<-\frac{L}{2} \qquad\quad\;\, {\rm (I)}\\
			0 & -\frac{L}{2}\le z \le\frac{L}{2}\quad {\rm (II)}\\
			\frac{1}{2}\omega_0^2\left(z-\frac{L}{2}\right)^2 & z>\frac{L}{2}\qquad\quad\;\; {\rm (III)},
		\end{cases}
		\label{eq:pot}
	\end{equation}
	with $\omega_0$ and $L$ real positive parameters. 
	We use atomic units throughout this work. For $L\to0$ this is the oscillator potential\cite{federico} $V(z)=\frac{1}{2}\omega_0^2z^2$. The larger $\omega_0$, the steeper the walls, and for $\omega_0\to\infty$, the system turns into 
	the infinite potential well 
	$ V_{\infty}(z)=V_0\theta\left(\left|z\right|-\frac{L}{2}\right)$ with $V_0\to\infty$.  The potential is depicted in the top panel of Fig.  \ref{fig:negative} for three values of $\alpha\equiv1/L^2\omega_0$, which is the adimensional parameter characterizing the problem, as explained below.
	
	The $j^{th}$ energy level $E(j;\omega_0,L)$ is our observable of interest $O(x;Q)$. 
	Let us first look at the exact problem. For $L\neq 0$ the  Schr\"odinger equation $\bigl[-\frac{1}{2}\frac{d^2}{dz^2}+V(z)\bigr]\psi(z)=E\psi(z)$ can be written 
	\begin{equation}
		\bigl[-\frac{1}{2}\frac{d^2}{dy^2}+v(y)\bigr]\phi(y)=\varepsilon\phi(y) \,,
		\label{eq:sch-scaled}
	\end{equation}
	with rescaled coordinate $y{\equiv}z/L$, rescaled energy  $\varepsilon{\equiv} L^2E$ and effective potential $v(y)=\left(y\pm\frac{1}{2}\right)^2/(2\alpha^2)$ and $v(y)=0$, for $|y|>\frac{1}{2}$ and $|y|<\frac{1}{2}$, respectively.  The parameter $\alpha=1/L^2\omega_0$ is a measure of the difference between the real system and  the infinite potential well, which vanishes for $\alpha\to 0$. 
	
	To obtain the exact solution, the Schr\"odinger equation is solved separately in each of the three regions (I), (II) and (III), and the solutions are glued by requiring the wavefunction and its first derivative to be continuous at the zone boundaries $y=\pm\frac{1}{2}$. This has to be done numerically; details can be found in App. \ref{sec:box}.
	The result is shown in the bottom panel of Fig. \ref{fig:negative} for the first energy levels $E(j;\omega_0,L)$ as function of $\alpha(\omega_0,L)$. For $L$ fixed and $\alpha\to0$, one obtains the eigenvalues of the infinite potential well $\mathcal E_j(L)=\pi^2j^2/(2L^2)$, whose separation increases linearly with $j$. For increasing $\alpha$, the eigenvalues are more and more evenly spaced, like for the harmonic oscillator. 
	
	Since the numerical solution is tedious, in practice one often uses perturbation theory to approximate the solutions of a  Schr\"odinger equation. This would correspond to evaluating $O^{\rm approx}(x;Q^R)$ using a perturbation approximation, taking this to be the final result, without  using it subsequently in \eqref{eq:inverse-approx} and \eqref{eq:connector-result}. We will call this in the following the \textit{direct approximation}. For this perturbation expansion we choose as zero order an infinite  well with width $L_0=L$.
	The expansion parameter is then $\alpha$, as defined above. The expansion to first order is carried out in App. \ref{sec:box}; the result is
	\begin{equation}
		%E^{approx}(j;\omega_0,L)
		E^{\rm approx}(j;\omega_0,L)=\mathcal E_j(L)\left[1-\frac{2\Gamma(\frac{1}{4})}{\Gamma(\frac{3}{4})}\sqrt{\alpha(\omega_0,L)}
		%\sqrt{\alpha(\omega_0,L)}
		%\frac{1}{\sqrt{L^2\omega_0}}
		\right]\,,
		\label{eq:2}
	\end{equation}
	with $\Gamma$ the Gamma function. 
	The first-order approximation has a very small computational cost compared to the exact numerical one.
	As Fig. \ref{fig:negative} shows, and as expected from perturbation theory, the approximate energy levels are close to the exact  $E(j;\omega_0,L)$ for small $\alpha$ and deviate for larger values, becoming unphysically negative above a critical threshold.
	
	The question is now whether COT, using the same approximation but combining it with knowledge from some model system, may yield better results. 
	It is intuitive to suppose that the effect of a finite curvature $\omega_0$ can be simulated by varying the width $L$ of an infinite potential well, i.e., to suppose that each level sees a surrounding with some effective width. As  model we will therefore choose  all possible  infinite potential wells, which means that
	we restrict the  model domain to potentials with $\alpha\to 0$.
	COT tells us to search for the effective width that best represents the real system, i.e., we will search for the connector $L^c$ for which $E(j;\omega_0,L) = \mathcal E_j(L^c_j)$\footnote{Note that here we connect the energy levels one by one, which is a choice. One would make this choice also to connect, for example, poles of Green's functions. \cite{Marco1}}.
	Note that here the infinite potential well plays a double role: it serves as model, and as the zero order of the expansion. This possible double role of a model in the case of a perturbative approximation is further discussed in \refsec{sec:pt}.

	To use COT we have to fulfill condition \textbf{[A]}: this is easily achieved for any potential with positive eigenvalues if we accept to have a different connector, i.e., a different effective width, for every energy level, since the energy levels in an infinite potential well can take any positive value.  
	The exact connector \eqref{eq:inverse} would be 
	$%L^c_j(L,\omega_0)
	L^c_{j\omega_0L}=\mathcal E^{-1}_{ j}(E(j;\omega_0,L))$. Since the energies $\mathcal E_{ j}(\tilde L)$ of an infinite potential well with width $\tilde L$ go as $1/{\tilde L}^2$, there is no unique inverse $\mathcal E^{-1}_{ j}(E)$.
	However, one of the two solutions of the resulting 
	quadratic equation for $L^c_j$ is negative and can be discarded by using positivity as exact constraint, so our example also fulfills condition \textbf{[B]}. 
	Finally, the exact connector reads:
	\begin{equation*}
		L^c_{j\omega_0L}=\frac{\pi j}{\sqrt{2E(j;\omega_0,L)}}.
		%{\sqrt{2E(j;\omega_0,L)}}.
	\end{equation*}
	As expected, to find the exact connector would require calculation of the exact energy levels $E$, which is what one wants to avoid. 
	Instead, the COT approximation strategy, required as condition \textbf{[C]}, is to  apply an equivalent approximation to both the real and the model system following \eqref{eq:inverse-approx}. Therefore, although the infinite potential well can easily be solved exactly, we   make the perturbation expansion also for the energy levels of the model. 
	The expansion of an infinite well with width $\tilde L$ around width $L$ reads: 
	\begin{align}
		\mathcal E_j^{\rm approx}(\tilde L)&=\mathcal E_j(L)+(\tilde L-L)\left.\frac{d\mathcal E_j(L')}{dL'}\right|_{L'=L}= \nonumber \\
		&=\mathcal E_j(L)\left[1-2\frac{\tilde L-L}{L}\right].
		\label{eq:appmodel}
	\end{align}
	Now we apply \eqref{eq:inverse-approx}, %of the main text, 
	i.e., we solve for the connector $L^{c,{\rm approx}}$ for which $E^{\rm approx}(j;\omega_0,L) = \mathcal E_j^{\rm approx}(L^{c,{\rm approx}})$ using Eqs. \eqref{eq:2} and \eqref{eq:appmodel}, which yields
	%which becomes
	\begin{equation}
		L^{c,{\rm approx}}_{\omega_0L}=L\left[1+\frac{\Gamma(\frac{1}{4})}{\Gamma(\frac{3}{4})}\sqrt{\alpha(\omega_0,L)}\right].
		\label{eq:conn1}
	\end{equation}
	Note that the model energy levels and first derivative are known as analytical functions, so there is no need to store and read data. Through $\alpha$, this approximate connector width $L^{c,{\rm approx}}$ depends on $L$ and $\omega_0$ but not on $j$: here, the connector is transferable, i.e., it could be calculated for one energy level and then used for all others. Such a transferability is not exact in general (see also the discussion in Subsec. \ref{subsec:53}), but could be used as an approximation, as we will illustrate below in the next example. Eq. \eqref{eq:conn1} also illustrates how the difference between $ L^{c,{\rm approx}}$ and $L$ becomes smaller and smaller as $\alpha$ tends to zero, namely as the real system approaches the model system itself.
	
	Once the connector has been obtained, the energy levels are taken from the exact solution of the model system following \eqref{eq:connector-result}: %of the main text:
	\begin{equation}
		%E^c(j;\omega_0,L) =
		E^c(j;\omega_0,L) = \mathcal E_j\bigl( L^{c,{\rm approx}}_{\omega_0L}\bigr)=\frac{\mathcal E_j(L)}{\left[1+\frac{\Gamma(\frac{1}{4})}{\Gamma(\frac{3}{4})}\sqrt{\alpha(\omega_0,L)}\right]^2}.
		\label{eq:3}
	\end{equation}
	%	Here, the first equality is what one would use in practice, when $\mathcal E_j$ has been tabulated. There may also be cases, like the present example, where the entry in the table can be converted to a simple expression, for example a rescaling, as expressed by the second equality. This, however, is not a key requirement for the connector to be efficient. Note that Eq. \eqref{eq:3} always leads to positive energies, contrary to the straightforward perturbation results Eq. \eqref{eq:2} that become unphysical for large $\alpha$, as discussed and shown in Fig. 1 in the main text.
	Note again that in the last step the \textit{exact} model function $\mathcal E_j$ is used, as prescribed by \eqref{eq:connector-result}.  The result \eqref{eq:3} is different from the direct approximation \eqref{eq:2}: the latter is the first-order expansion of \eqref{eq:3}. Using the COT expression \eqref{eq:3} does not add computational cost, but results are improved significantly, as shown in 
	%Results are shown in Fig. \ref{fig:negative}. 
	%Contrary to $E^{\rm approx}_n$, 
	Fig. \ref{fig:negative}. They are always better  than the direct approximation $E^{\rm approx}(j)$, and they remain physical over the whole parameter range, which reflects an implicit approximate resummation of perturbation theory to infinite order through the use of the exact model $\mathcal E_j$; see \refsec{sec:pt} for a general discussion of this point. The connector $L^c_j$ formalizes the intuitive interpretation of an effective width that determines the energy levels in the real system and shows that this effective width is approximately independent of $j$. This result indicates a promising route to obtain quick estimates for energy levels.

	\section{Quick estimates from a connector approximation }
	
	The toy example of the one-dimensional potential indicates that connector approximations may be used to speed up calculations or, equivalently, to improve the results of rough approximations without additional computational cost. In the following, we will examine this route by looking at  a more realistic situation, using as example the calculation of a band structure. Here, the focus will not be to improve the accuracy, e.g., by simulating many-body effects, but to obtain quick results for a given effective hamiltonian.
	
	\subsection{Connector approximation to calculate an independent-particle band structure}
	
	The perturbative connector approximation suggested by the simple example above could in principle be extended to real systems. Here, we will investigate another route, since 
	it is important to note that COT approximations are not limited to perturbation theory. We will therefore examine energy levels for real materials, by changing the approximation: inspired by the perfectly transferable connector of above, we will use transferability as approximation.
	
	As illustration, we will apply COT to obtain in an efficient way the KS band structure of solid cubic He \cite{Schuch1961}, with given KS potential $V({\bf r})$. For a material with a periodic  potential the HEG is the simplest possible model. 
	In the HEG,  the band structure is the free electron parabola folded into the Brillouin zone, and its potential $V^c$ is constant in space, characterized by one number $\mathcal V^c$  {$\in \mathbb{R}$} with respect to the chemical potential. This potential allows us to lower or raise the HEG energy levels at will - so condition \textbf{[A]} is satisfied -, and it will play the role of the connector.

	For many materials it may be difficult to find a one-to-one relation between their bands and the  HEG ones. Therefore, 
	we do not connect bands but derive the band structure from the spectral function, which is
	the trace over the imaginary part of the KS Green's function $G({\bf k},\omega)$. To be precise, we impose 
	the connector equality Eq. (\ref{eq:final})  as 
	\begin{equation}
		G_{{\bf K}{\bf K}}({\bf k},\omega;[V]) = \mathcal{G}_{{\bf k}\omega{\bf K}}(\mathcal{V}^c_{{\bf k}\omega{\bf K}}),
		\label{eq:supmat-G}
	\end{equation}
	where $G$ is the KS Green's functions of the real material, and ${\bf K}$ are reciprocal lattice vectors.  The Green's function of the HEG is diagonal in ${\bf K}$, so we can represent it by $\mathcal G_{{\bf k}\omega{\bf K}}$, which is a function of ${\bf k}+{\bf K}$. It is also a function of the value $\mathcal{V}^c$ of the homogeneous potential,
	\begin{equation}
		\mathcal{G}_{{\bf k}\omega{\bf K}}(\mathcal{V}^c) = \frac{1}{\omega - |{\bf k}+{\bf K}|^2/2 - \mathcal{V}^c}\,,
		\label{eq:model-heg-g}
	\end{equation}
	where the frequency $\omega$ includes an infinitesimal imaginary part.
	This expression is readily calculated as function of $(\omega-\mathcal{V}^c)$ and $|{\bf k}+{\bf K}|$ and could be either stored or recalculated on the fly.
	From Eq. \eqref{eq:supmat-G}, the exact connector potential corresponding to \eqref{eq:inverse} is
	\begin{equation}
		\mathcal{V}^c_{{\bf k}\omega{\bf K}} = \omega - \frac{|{\bf k}+{\bf K}|^2}{2} -\frac{1}{G_{{\bf K}{\bf K}}({\bf k},\omega;[V])}.
		\label{eq:exact-Vc}
	\end{equation}
	It is important to note that for our purpose it is enough to ask for the diagonal $G_{{\bf K}{\bf K}}$: the HEG would not be able to simulate off-diagonal elements, at least not in a physically reasonable way, and therefore asking for off-diagonal elements would violate condition \textbf{[A]}. 
	
	As one can see, for the
	band structure the wavevector ${\bf k}$ plays the role of $x$, and the
	connector potential is in principle different for each wavevector ${\bf k}$. Moreover, it should depend on the band;
	here, this is translated into a $\omega$-dependence.
	
	Strictly speaking, since the spectral function is the imaginary part of the trace (here, a sum over ${\bf K}$), we could have asked for 
	$\sum_{\bf K}{\rm Im}\,G_{{\bf K}{\bf K}}({\bf k},\omega;[V]) = \sum_{\bf K}{\rm Im\,}\mathcal{G}_{{\bf k}\omega{\bf K}}(\mathcal{V}^c_{{\bf k}\omega})$, with a ${\bf K}$-independent connector potential $\mathcal{V}^c_{{\bf k}\omega}$. However, to invert this relation and select a unique answer would be much more cumbersome than to invert Eq. (\ref{eq:supmat-G}), for which condition \textbf{[B]} is easily satisfied.
	
\begin{figure}
		\centering
		\includegraphics[width=0.48\columnwidth]{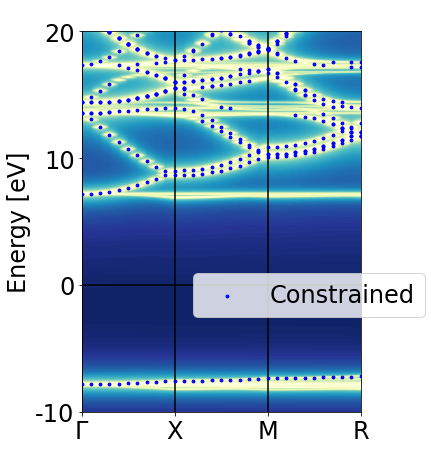}
		\includegraphics[width=0.455\columnwidth]{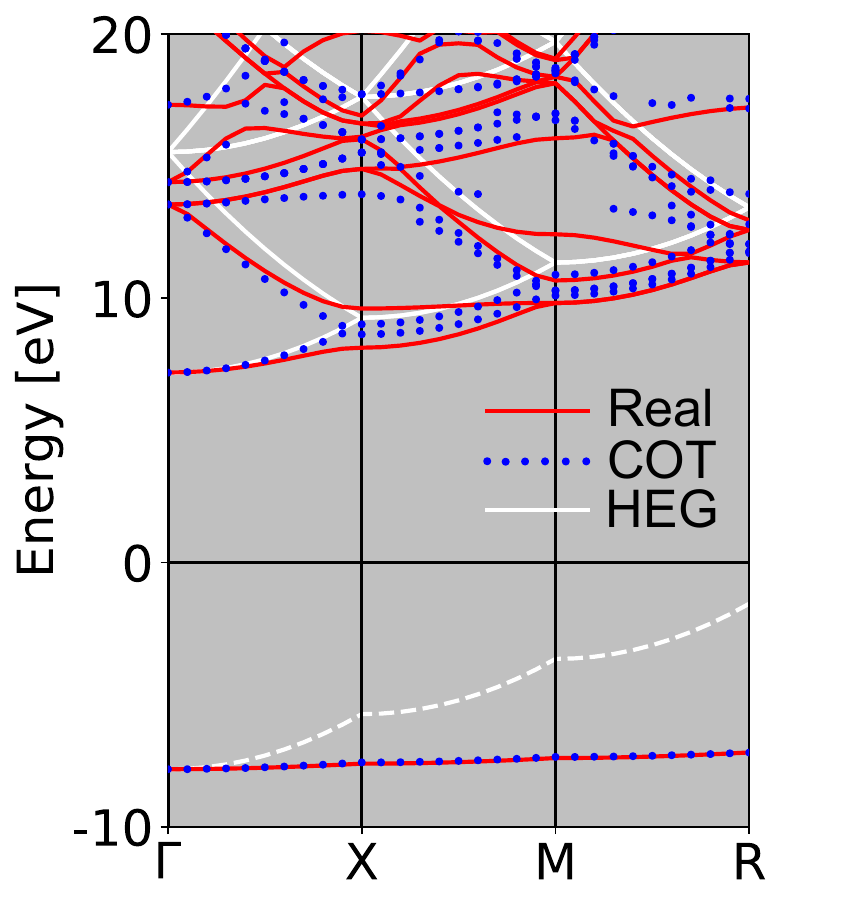}
		\caption{Band structure of cubic helium, with the energies aligned to the Fermi energy. Left: connector result, plain (intensity plot) and constrained (blue dots), following the prescription in the text. Right: comparison of first principles results (red line) with constrained connector (blue dots). In white, HEG folded into the Brillouin zone of helium and attached to the second band at $\Gamma$. For comparison of the lowest band, which corresponds to a localized state,  the lowest HEG band is used.}
		\label{fig:BS}
	\end{figure}

	The exact connector at a point ${\bf k}_0$ can be found by calculating $G_{{\bf K}{\bf K}}({\bf k}_0,\omega;[V])$ and using \eqref{eq:exact-Vc}. To get the band structure, one should do this for every ${\bf k}$, which would again be useless, as it would be equivalent to a full solution of the original problem. We therefore move on to our approximation. As anticipated, it will consist in supposing 
	approximate \textit{transferability} of the connector through  cancellations in Eq. (\ref{eq:inverse-approx}). Our approximation is hence to
	neglect the ${\bf k}$-dependence of the connector by using at all ${\bf k}$ its value at ${\bf k}_0$. 
	Of course, applied directly to the band structure or spectral function, transferability would be a bad approximation, since the direct approximation would read $G_{{\bf K}{\bf K}}({\bf k},\omega;[V])\approx G_{{\bf K}{\bf K}}({\bf k}_0,\omega;[V]) $ and therefore yield only flat bands. Also in the COT scheme, we diagonalize the KS hamiltonian only once, at ${\bf k}_0=\Gamma$, but now this yields the exact $\mathcal{V}^c_{\Gamma\omega{\bf K}}$, and we then approximate 
	\begin{equation}
		\mathcal{V}^{c\,approx}_{{\bf k}\omega{\bf K}} = \mathcal{V}^c_{{\bf k}_0\omega{\bf K}} \,,
	\end{equation}
	according to the strategy required by condition \textbf{[C]}.
	Following \eqref{eq:connector-result}, this yields the band structure at all ${\bf k}$-points as
	\begin{equation}
		G^{c\,approx}_{{\bf K}{\bf K}}({\bf k},\omega;[V]) = \frac{1}{ - \frac{|{\bf k}+{\bf K}|^2}{2}  + \frac{|{\bf k}_0+{\bf K}|^2}{2} +\frac{1}{G_{{\bf K}{\bf K}}({\bf k}_0,\omega;[V])}}\,.
	\end{equation}
	This corresponds to a diagonal approximation of the in principle exact Dyson equation that relates $G_{\bf k}$ and $G_{{\bf k}_0}$. The result of this straightforward COT approximation for the case of  solid cubic He \cite{Schuch1961} is shown in the left panel of Fig. \ref{fig:BS} (see footnote \footnote{The reference band structures of cubic helium and hexagonal hydrogen, with experimental lattice parameters\cite{Schuch1961,Ishmaev1983}, %4.242 $\AA$ and 3.704 $\AA$ respectively, 
		have been obtained from the diagonalization of the KS-LDA Hamiltonian of size 5.1 Ha and 3.4 Ha in reciprocal space, respectively. The Hamiltonian has been calculated 
		using local Troullier-Martins\cite{Troullier1993} pseudopotentials in the Abinit code\cite{abinit-code} with $\Gamma$-centered  5$\times$5$\times$5 and 3$\times$3$\times$3 $\bfk$-point grids, respectively.} for computational details).
	
	The straightforward connector result is blurred and suggests spurious energy levels, since it violates an exact constraint, namely, that the poles of the Green's function should not depend on ${\bf K}$. Still,  one can  identify bands approximately. We can therefore  impose the exact constraint by choosing for each group of poles the one with the largest oscillator strength: this reflects the situation where the real system and the HEG are most similar, which means, where COT should work best. Details of how this is done in practice are given in App. \ref{sec:app-poles}. The result is given by the blue dots in the left panel of Fig. \ref{fig:BS}: the constrained connector makes the bands sharp and eliminates some spurious non-dispersing structures. 
	
	The right panel of Fig. \ref{fig:BS} compares the \textit{ab initio} and the constrained connector results. They are impressively similar, knowing that only one diagonalization, at $\Gamma$, is needed. The rest of the information is imported from the HEG. The HEG bands, aligned to the second valence band at $\Gamma$, are the white lines for comparison. Of course, the higher bands of He have some resemblance to free electrons, but the differences are significant and overall well reproduced by the connector. The lowest valence band is due to localized $s$-states and far from the additional free electron band given by the dashed white line aligned at $\Gamma$. The connector correctly predicts the almost complete absence of dispersion for this highly localized state, which is very far from a HEG or simple metal. 
	
\begin{figure}
		\includegraphics[width=.5\columnwidth]{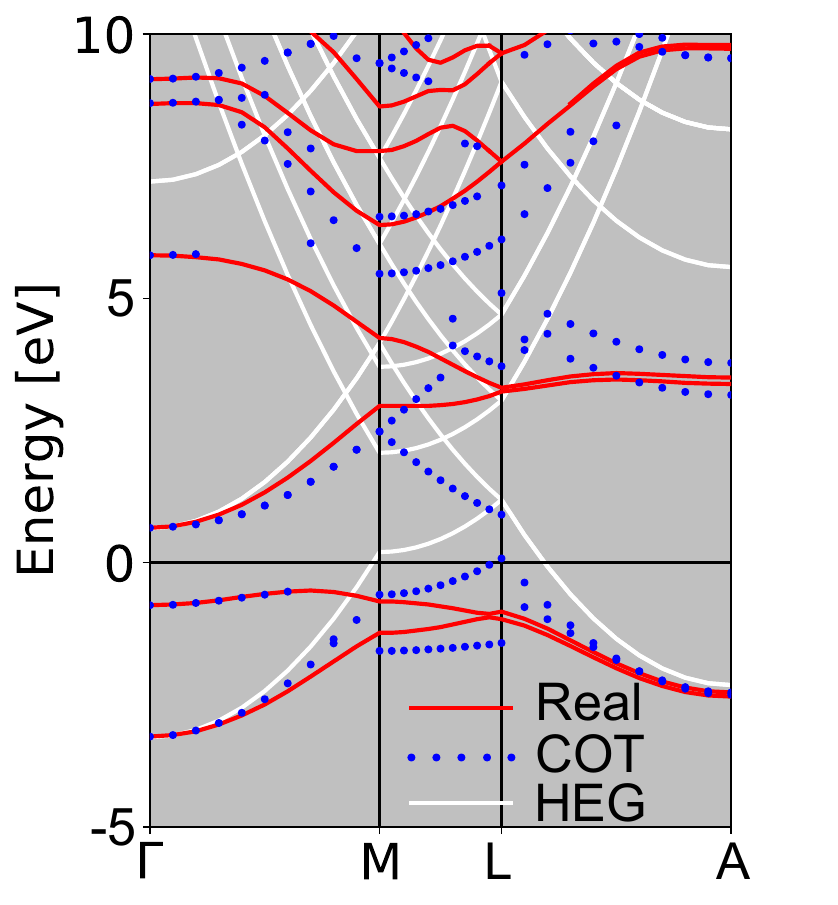}
		\caption{Band structure of solid hydrogen, aligned to the Fermi energy. Reference result (red continuous) and constrained connector (blue dots). In white, HEG folded into the Brillouin zone of hydrogen and attached to the first valence band at $\Gamma$.}
		\label{fig:BS-sup-H}
	\end{figure}

	By construction, the further the real system is from the model, the less well will a simple approximation to COT perform. We also want to illustrate this case, so Fig.  \ref{fig:BS-sup-H} displays constrained connector results for solid hydrogen\cite{Ishmaev1983}, obtained in the same way as for the example of helium.  As the figure shows, for this material the \textit{ab initio} band structure is very far from the HEG one. The main deviations from the HEG are well described by COT, for example, the existence of a double band below 5 eV between L and A, or of the second band between $\Gamma$ and M. At the same time, supplementing the information at $\Gamma$ with the HEG yields a realistic idea of the dispersion of the bands, which would be completely flat, had we applied our approximation directly to the calculation of $G$, without using the connector. The quality of the connector band structure is instead less satisfactory between M and L, as one would expect for a path far from the reference point, i.e. $\Gamma$. To improve, one could produce the exact results in a few more ${\bf k}$-points, assuming transferability of the connector over a part of the Brillouin zone only, or one may go beyond the HEG, e.g. moving to a periodic model, which is more pertinent in the spirit of COT. This will be left for future work.
	In its present approximation, the connector approach may not replace a full band structure calculation for all materials, but it is certainly a very fast way to already obtain much insight. 
	
	\section{Connector approach to the design of functionals}
	
	The potential use of connector approximations is not limited to speeding up calculations. Indeed,  when the parameters $Q^R$ that characterize the real system constitute a function, i.e., when $O$ is functional of $Q$, \eqref{eq:connector-result} together with \eqref{eq:inverse-approx} stand for an approximate functional. This functional benefits from pre-calculated ingredients that are universal, in the sense that once the model is decided, they will be the same for all real systems.  The design of functionals is therefore a very promising field of applications for COT, as we will illustrate in the following. 
	
	\subsection{Exchange-correlation potential: density functional from a connector approximation}
	One of the prototype problems for the design of functionals is the search for a simple but reliable  Kohn-Sham exchange-correlation potential  $v_{\rm xc}({\bf r};[n])$  beyond the LDA, which is still a subject of intense research (see e.g.  \cite{Gunnarsson1977,Alonso1977,Alonso1978,Gunnarsson1979,Gunnarsson1980,Cuevas2012,LangrethPerdew1980,PerdewWang1986,PBE1996,Perdew1999,Perdew2008,Sun2015,Burke2012,Cohen2012,Becke2014,Pribram-Jones2015,Jones2015}). Our aim here is to demonstrate that COT has the potential to suggest new density functionals, and to explain in which way they lead to an improvement with respect to the LDA.  
	
	To apply COT, for close comparison with the LDA we use again the HEG as model system. Contrary to above in the example of the band structure, now the HEG will not be defined by its potential, but by its density, in other words, the connector will be a number giving an effective homogeneous density.
	The exact connector condition \eqref{eq:final} reads:  
	\begin{equation} %\label{exact_connection}
		v_{\rm xc}(\mathbf{r},[n]) = v_{\rm xc}^h(n^{c}_{\mathbf{r}}), 
		\label{eq:vxc-exact}
	\end{equation}
	where $n^{c}_{{\bf \mathbf{r}}}$ is the unknown homogeneous density that yields the exact value of the xc potential $v_{\rm xc}(\mathbf{r},[n])$ at position $\mathbf{r}$, when used to evaluate the xc potential of the HEG $v_{\rm xc}^h(n^h)$. 
	Note that $n^{c}_{\mathbf{r}}$ is a homogeneous density, but a different $n^{c}_{\mathbf{r}}$ is used for the potential in every point $\mathbf{r}$. This freedom is crucial, because otherwise the exact condition (\ref{eq:vxc-exact}), which reflects condition \textbf{[A]}, could not be fulfilled. In this way, instead, all $v_{\rm xc}$ of real systems can be reached as long as they are negative, as the HEG xc potential $v^h_{\rm xc}$ spans all negative values.\footnote{It should be noted that 
		the condition will often be fulfilled straight away, but there are exceptions. In particular, if one is interested, for example, in the dissociation of linear molecules, one must be aware of the fact that the xc potential $v_{\rm xc}({\bf r})$ of these one-dimensional systems shows  positive bumps, see, e.g., \cite{Perdew1985a,Buijse1989,baerends-bump,Tempel2009,Helbig2009,Wetherell2019} whereas the exchange-correlation potential in the HEG is always negative. Therefore, if one wants to connect these systems to the HEG  additional action, such as an extension of the domain, would be needed to fulfill condition \textbf{[A]}. 
		%\textcolor{red}{do we also have a reference for a positive $v_{\rm xc}$?} \textcolor{magenta}{Dissociation of molecules (toy models): see e.g. Fig 1 in \url{https://arxiv.org/pdf/1812.02661.pdf}; see Figs. 3-4 in \url{https://arxiv.org/pdf/0812.1247.pdf} }
		%\textcolor{magenta}{See \cite{Levy1985} and refs therein for atoms!}
		This, however, goes beyond the scope of the present illustration.} Concerning condition \textbf{[B]} there is no problem, since $v_{\rm xc}^h$ is a monotonous function that can be inverted straightforwardly.
	
	To illustrate that we can also fulfill \textbf{[C]} and obtain efficient expressions for practical use, our scope is now to find an approximate connector, for which
	\eqref{eq:inverse-approx} reads 
	\begin{equation}
		n^{c,{\rm approx}}_{\bf r} = (v_{\rm xc}^{h,{\rm approx}})^{-1}(v_{\rm xc}^{{\rm approx}}({\bf r};[n]))\,.
		\label{eq:con-nrc}
	\end{equation}
	
	A first crude approximation could be to cut the Coulomb interaction $1/|{\bf r}|$ beyond a short distance $r=r_c$, without renormalizing the remaining short-range part of the interaction. As $v_{\rm xc}$ entirely stems from the Coulomb interaction, dropping a substantial part of the interaction decreases the exchange-correlation potential, and even makes it vanish with the cutoff radius $r_c\to 0$.  
	The connector result corresponding to the very same approximation, instead, should  tend to the LDA, because now we do not apply the approximation to the potential, but only to the way the potential in a given point ``sees'' the environment, limiting it, for $r_c\to 0$, to the environment close to the point ${\bf r}$ where the potential is calculated. We can therefore expect a much more meaningful result.
	
	\subsection{Benchmarking the description of non-locality}
	In order to illustrate this and the following discussion, we cannot compare to the exact functional $v_{\rm xc}$, which is unknown.  Instead, we take an established non-local functional as target, and we will do all approximations and calculations consistently with respect to this functional. It is important to note that the aim is \textit{not} to elucidate the quality of this target functional, but to use it as benchmark for the connector approach, and in particular, how well it can capture non-local effects beyond the LDA.
	
	We choose as  
	target functional an expression based on a weighted density approximation (WDA) of the xc hole $n_{\rm xc}$ \cite{Gunnarsson1977,Alonso1977,Alonso1978,Gunnarsson1979}, with the weight function proposed in \cite{Gunnarsson1980}. The xc energy reads 
	\begin{equation}
		E_{\rm xc}^{\rm WDA}[n] = \int d \mathbf{r}  d\mathbf{r'} \frac{n(\mathbf{r})n(\mathbf{r'})}{2\mathbf{|r-r'|}} C(\tilde{n}(\mathbf{r,r'})) \Big(  1- e^{\frac{- \lambda(\tilde{n}(\mathbf{r,r'}))}{\mathbf{|r-r'|^5}}} \Big),
		\label{eq:E-WDA}
	\end{equation}
	with $\tilde{n}({\bf r}, {\bf r'})= [n({\bf r}) + n({\bf r'})]/2$ \cite{Pablo2000}  and the Coulomb interaction $1/|{\bf r}-{\bf r}'|$ appearing at two places, as factor in the integral and in the exponent to the fifth power. The functions $C$ and $\lambda$ are set by the sum rule $\int d {\bf r'} n_{\rm xc}({\bf r}, {\bf r'- r})=-1$, and by making
	%, that depend on $\tilde n$, 
	$E_{\rm xc}$ exact in the HEG. The functional derivative of $E_{\rm xc}^{\rm WDA}$ with respect to the density yields the exchange-correlation potential $v_{\rm xc}^{\rm WDA}$ given in the appendix [Eq. \eqref{eq:vxc-wda}], together with the functions $C$ and $\lambda$ [Eq. \eqref{eq:constants}]. This potential is our target. To be precise, we will not investigate the density that is created by the potential, but the potential itself, for a given input density. The resulting density, which requires the use of the potential in a self-consistent cycle, would be an important topic on itself \cite{Kim2013,Wasserman2017,Vuckovic2019}. 
	
\begin{figure}[t]
		\includegraphics[width=
		\columnwidth]{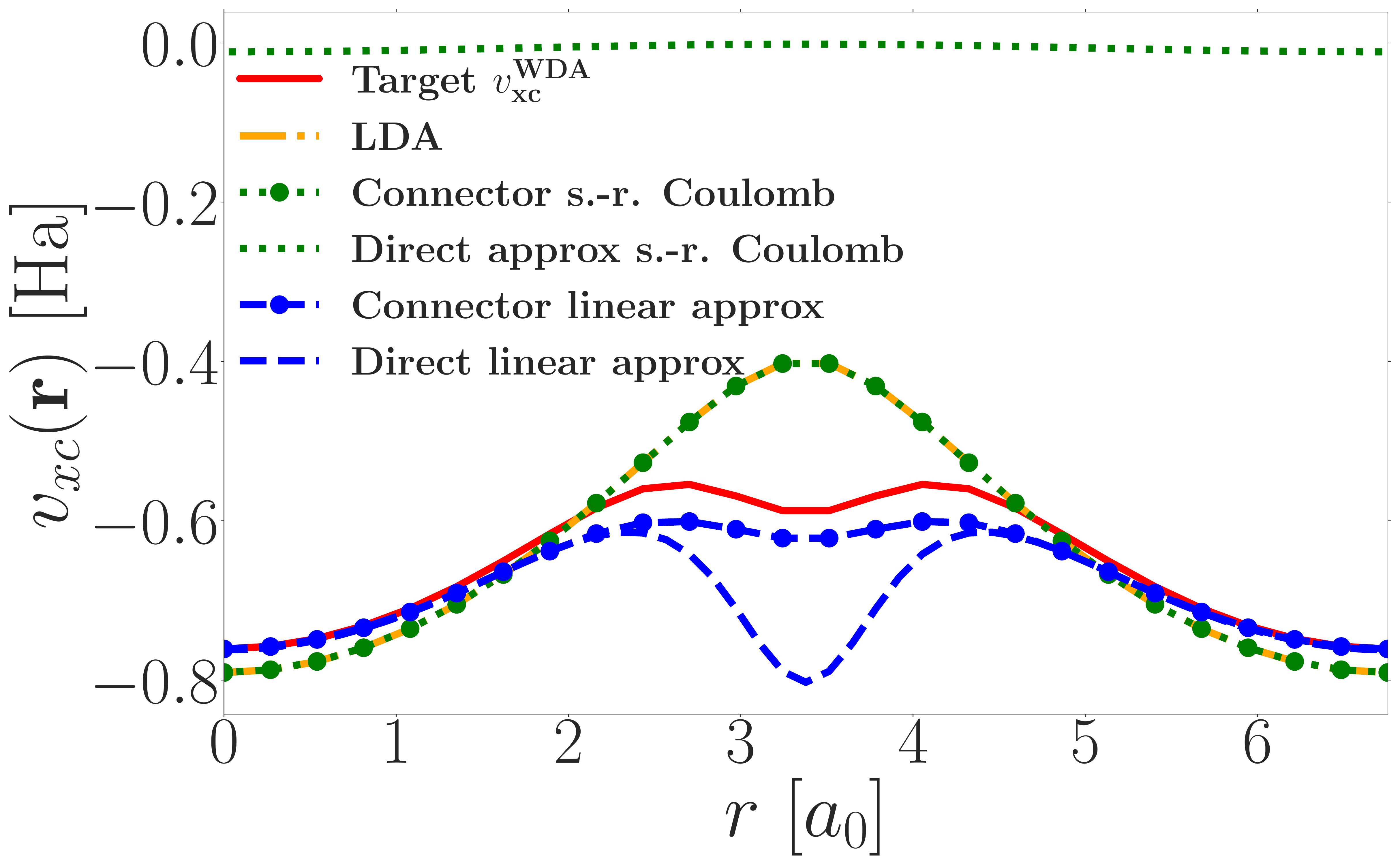}
		\caption{Target $v_{\rm xc}^{\rm WDA}({\bf r})$ (red line), and approximations, for a system with periodic density. Coulomb cutoff at short-range (s.-r.): direct approximation (dotted green) and used in COT (dotted green with filled circles); linear expansion: direct approximation (blue dashed) and used in COT (blue dashed with filled circles); LDA (yellow line).
			Minimum and maximum densities are  $0.0402 \ a_0^{-3}$ and $0.3776 \ a_0^{-3} $, reciprocal lattice vector $a=0.93 \ a_0^{-1}$.     
		}
		\label{fig:sf}
	\end{figure}

	The target potential   $v_{\rm xc}^{\rm WDA}$ is shown in Fig. \ref{fig:sf},
	for an inhomogeneous ``real system'' with periodic density $n(\mathbf{r})= A \cos (\mathbf{a}\cdot\mathbf{r}) + B $, reciprocal lattice vector $\mathbf{a}$ and parameters $A$ and $B$. Here ${\bf a}$ and the maximum density correspond to the case of solid argon. The potential is negative for all systems examined in our work, so it can in principle be reproduced by $v_{\rm xc}^h$ as required by condition \textbf{[A]}. 
	The figure also shows the results of the  crude approximation obtained when the Coulomb interaction  $1/|{\bf r}-{\bf r}'|$ in \eqref{eq:E-WDA} is  cut beyond a very short distance $r=r_c$, which limits the integral boundaries. Applied directly to  $v_{\rm xc}^{\rm WDA}$, this yields, as anticipated above and as one would expect, an almost vanishing xc potential, here obtained with $r_c=0.1 \ a_0$ with $a_0$ the Bohr radius, less than 2\% of the unit cell length. The connector result based on the very same approximation
	is now obtained by using the HEG xc potential to simulate the real system, \textit{via}
	\begin{equation} %\label{connector Vxc}
		v_{\rm xc}^c(\mathbf{r},[n]) = v_{\rm xc}^h\left(n^h=n_{\mathbf{r}}^{c,{\rm approx}}([n])\right)\,,
		\label{eq:vxc-approx-gen}
	\end{equation}
	which is the equivalent of \eqref{eq:connector-result}. As the figure shows, this potential does not tend to zero, but to the LDA. 
	The reason is that the $v_{\rm xc}$ of both the real system and the HEG tend to zero in  \eqref{eq:con-nrc}, but the connector construction leads to strong error cancelling. 
	The huge improvement of the final result with respect to the direct approximation  makes the LDA a prototype illustration for the fulfillment of condition \textbf{[C]} and for the power of the connector approach. 
	
	Still, the LDA has significant errors especially in regions of low density, because the approximation is too crude. In principle one could remedy by increasing the range of the Coulomb interaction, going systematically towards the exact result.  However, this approximation would be of limited practical interest. 
	
	An alternative and more useful
	%is $v_{\rm xc}({\bf r},[n]) = v_{\rm xc}^h(n^c_{h{\bf r}}) $, where $h$ indicates ``homogeneous''and the subscript ${\bf r}$ stresses the fact that we obtain a different homogeneous connector density for every point ${\bf r}$ of the real system. 
	approximation, which also fits in our strategy required by condition \textbf{[C]}, is a first-order expansion around  
	%The LDA approximates $n^{c,\rm approx}_{h{\bf r}}([n]) = n({\bf r})$.
	%Starting from the exact expression allows us to go beyond the LDA in a systematic way. One possibility to obtain a highly nonlocal density functional is to expand the real and model potentials 
	%around 
	a given homogeneous density $n_0$ \cite{DFT1,Palummo1999}. We will now use this approximation in the connector scheme to derive a new functional. We will then compare the result to the LDA, and benchmark its performance again by examining how well it can reproduce non-local effects, with the WDA defined as target.
	
	Using a first-order expansion as approximation in \refeq{eq:final} yields: 
	\begin{eqnarray}\label{first_order_expansion} 
		v_{\rm xc}^h(n_0)+
		\int d \mathbf{r}'  (n(\mathbf{r}')-n_0)f_{\rm xc}(|\mathbf{r}-\mathbf{r}'|;n_0)  =  v_{\rm xc}^h(n_0)+  (
		n^{c,{\rm approx}}_{\mathbf{r}} -n_0)f^h_{\rm xc}(n_0) , 
	\end{eqnarray}
	where $f_{\rm xc}(|\mathbf{r}-\mathbf{r}'|;n_0)=\left.\delta v_{\rm xc}(\mathbf{r},[ n])/\delta n(\mathbf{r}')\right|_{n=n_0}$ is the static nonlocal xc kernel of the HEG with density $n_0$, and $f^h_{\rm xc}(n_0)=\left.d v_{\rm xc}^h(n_h)/d n_h\right|_{n_h=n_0}$ is its limit of zero wavevector, which corresponds to the case where variations are restricted such that the density remains homogeneous, i.e., we remain within the parameter space given by the model. The zeroth order term $v_{\rm xc}^h(n_0) $ on both sides cancels. Finally, inverting the equation we get  the approximate connector  
	\begin{equation}%\label{connector 1st order} 
		n^{c,{\rm approx}}_{\mathbf{r}}([n])= \frac {1}{f_{\rm xc}^h(n_0)} \int d\mathbf{r}' \ 
		n(\mathbf{r}') f_{\rm xc}(|\mathbf{r}-\mathbf{r}'|;n_0)\,,
		\label{vxc_conn}
	\end{equation}
	which should be used to reproduce the quantity of interest $v_{xc}$ according to the prescription of \eqref{eq:vxc-approx-gen}.
	For a density of the real system that varies slowly on the scale of spatial decay of $f_{\rm xc}(|\mathbf{r}-\mathbf{r}'|)$ the connector tends to $n^{c,\rm approx}_{{\bf r}}([n]) = n({\bf r})$, the LDA,
	which is exact in this limit. For a very quickly varying periodic density we obtain the mean density, as one would expect, and which is instead completely missed by the LDA. The approximate connector interpolates between these two limits: it displays the degree of nearsightedness of the xc potential, which is obtained from \eqref{eq:vxc-approx-gen}.
	For the exact $f_{\rm xc}$ of the HEG one can use parametrized data. \cite{Corradini1998,Moroni1995} This means that, similarly to the LDA, one can in principle benefit from high-level calculations performed once and for all in the model. 
	
	Up to here, the derivation and discussion of this linear-expansion connector approximated $v_{\rm xc}$ was general. In the following, 
	we again want to test this COT approximation by reproducing $v_{\rm xc}^{\rm WDA}$.
	In order to simulate a realistic situation, $f_{\rm xc}$ has to be consistent with the $v_{\rm xc}$ that one is heading for. Therefore, for this test we should not use expressions for $f_{\rm xc}$ given in the literature, but
	we have to calculate the functional derivative of our target xc potential, and evaluate it at the constant density $n_0$. 
	The resulting $f_{\rm xc}^{\rm WDA}$ can be found in the Appendix, \refeq{eq:fxc-wda}. 
	
	The linear approximation is not yet well defined, because it depends on the homogeneous density $n_0$ around which we expand. This density may be chosen to be a different one for every point ${\bf r}$ where we calculate the potential.  
	To focus 
	on the improvement with respect to the LDA, we choose for $n_0$ the local density $n({\bf r})$, i.e. the direct approximation is a first-order expansion around the LDA. 
	
	\begin{figure}[t]
		\includegraphics[width=
		\columnwidth]{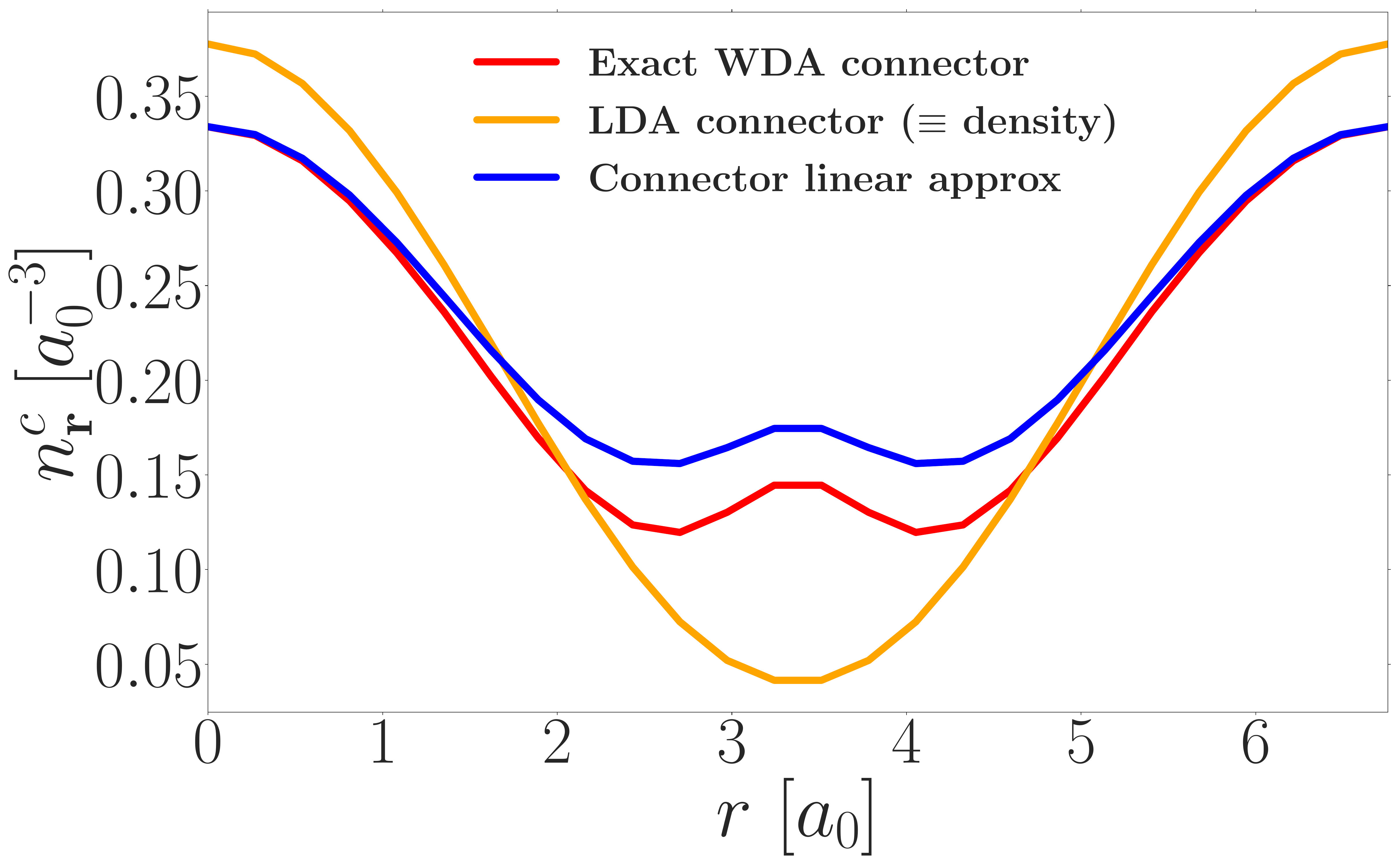}
		\caption{Comparison of connector densities: exact WDA connector $n_{\bf r}^{c,WDA}$  (red line), connector based on a linear expansion in the density as given by \refeq{vxc_conn} (blue  line), and the density itself, corresponding to the local density approximation connector (yellow line), for the periodic system of Fig. \ref{fig:sf}.}
		\label{fig:exactcon}
	\end{figure}
	
	Let us first look at the performance of the linear approximation itself, used directly on the potential without going through the connector scheme. The comparison in Fig. \ref{fig:sf} shows that the WDA potential is well described by the direct linear approximation in regions of high density (large $|v_{\rm xc}|$), but not where the density is low. We now move on to COT, and first examine the connector density $n^c_{\bf r}$ given by \eqref{vxc_conn}
	and shown in Fig. \ref{fig:exactcon} for the same periodic system as above. 
	The connector density that yields the LDA is simply the  density itself, which has its minimum in the middle of the unit cell. The exact WDA connector $ n^{c,WDA}_{\bf r} = (v_{\rm xc}^{h})^{-1}(v_{\rm xc}^{WDA}({\bf r};[n]))$ spans a smaller range of values than the local density, confirming that it represents an effective density that is suitably averaged over a range around the local density. It is far from trivial, even in this simple system: where the density is very low, the exact connector shows a bump, which is a feature that would not be easy to guess. One may explain it with the fact that in regions of low density the system is more far-sighted, which, in the present system, mixes in more of the higher densities, and with the non-monotonous distance dependence of effective interactions, such as $f_{\rm xc}$. The approximate connector density captures this effect very well, with a remaining discrepancy that is much smaller than the difference to the LDA.

	Finally, Fig. \ref{fig:sf} shows the exchange-correlation potential resulting from our connector. It gives by far the best result of all approximations, illustrating the fact that the connector \eqref{vxc_conn}
	takes into account a significant amount of non-locality. It shows that, by using the very same linear expansion through COT, strong improvement is obtained, without additional cost.

\begin{figure}[!h]
		\includegraphics[width=\columnwidth]{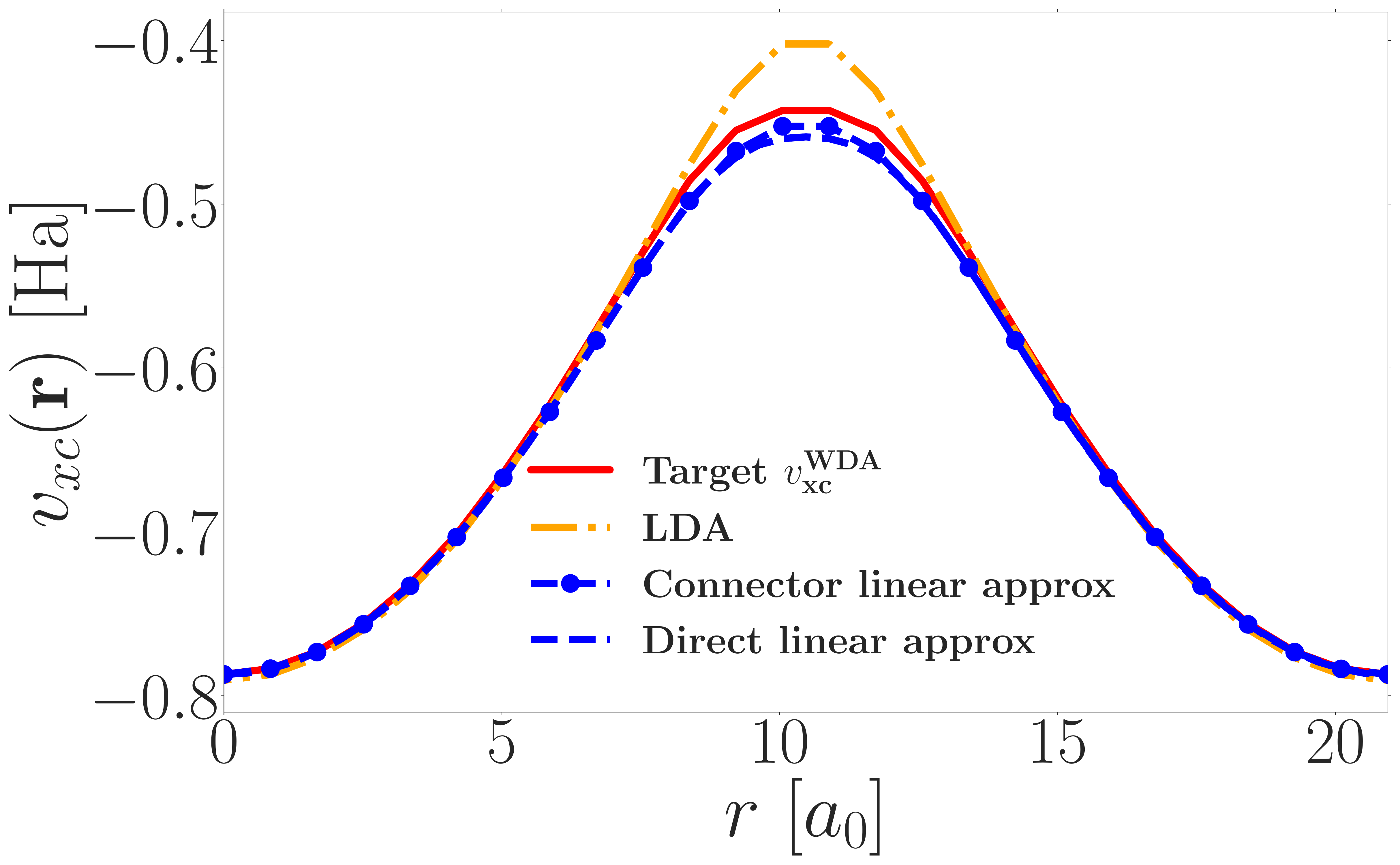}
		\caption{Comparison between the target WDA xc potential (red line) and different approximations: LDA (yellow line), direct linear approximation (blue dashed) and using the connector (blue dashed with filled cycles)
			for slowly varying density. In this case $n(\mathbf{r})=A \cos(\mathbf{a}\cdot\mathbf{r}) +B $, where $a= 0.3 \ a_0^{-1}$, $n_{max}= 0.3776 \ a_0^{-3}$ and $n_{min}= 0.0402 \ a_0^{-3}$.}
		\label{fig:ldalimit}
	\end{figure}

	Since the connector tends to the LDA for slowly varying density,  the improvement  with respect to LDA is particularly important when the density is quickly varying; however, some improvement is also found in the case of more slowly varying density, i.e. smaller $\mathbf{a}$, as shown in Fig. \ref{fig:ldalimit}. The figure also illustrates how the connector result tends to the LDA, which itself tends to the exact result, when the density approaches the homogeneous limit.
	
	\subsection{A connector adapted to each object of interest}
	\label{subsec:53}

	The above results confirm that COT is a promising approach to design new simple functionals capable to improve results with respect to the LDA. It was our choice to concentrate on $v_{\rm xc}$ for this illustration, because it is the unknown ingredient of the Kohn-Sham potential. Of course, other choices for illustrations would have been possible: we could have examined the exchange-correlation energy or the xc energy density $\epsilon_{\rm xc}({\bf r})$, or we could have looked at single pieces, such as $v_{\rm x}({\bf r})$ and $v_c({\bf r})$, or at the correlation contribution to the kinetic energy alone. It is important to understand what such a choice implies and how the different approaches are linked within the COT framework.
	
	Let us start with the question what happens if we separate $v_{\rm xc}({\bf r})$ into two contributions, e.g., $v_{\rm x}({\bf r})+v_{\rm c}({\bf r})$. Suppose that for the systems of interest condition \textbf{[A]} can be fulfilled, e.g., an exact connector exists in principle, for the sum as well as for both pieces. This would mean that
	\begin{eqnarray}
		v_{\rm xc}({\bf r}) &=& v^h_{\rm xc}(n^c_{\bf r})\nonumber\\
		v_{\rm x}({\bf r}) &=& v^h_{\rm x}(n^c_{{\bf r},{\rm x}})\nonumber\\
		v_{\rm c}({\bf r}) &=& v^h_{\rm c}(n^c_{{\bf r},{\rm c}})\,:
	\end{eqnarray}
	in principle, each contribution is calculated with its own connector, and there is no reason to suppose the connector that reproduces exactly the exchange contribution, 
	$ n^{c}_{{\bf r},{\rm x}} = (v_{\rm x}^{h})^{-1}(v_{\rm x}({\bf r};[n]))$, to be the same as the exact connector for the correlation contribution $ n^{c}_{{\bf r},{\rm c}} = (v_{\rm c}^{h})^{-1}(v_{\rm c}({\bf r};[n]))$, nor to suppose that it equals $n^{c}_{{\bf r}}$. 
	As a consequence, even when a consistent approximation is chosen for all pieces, most often also the approximate connectors will be different.

	This is a general point concerning the separation of the object of interest into two or more contributions, not limited to the separation illustrated here. %\textcolor{red}{\textbf{skip rest of sentence}}, nor to $v_{\rm xc}$. 
	Such a separation may be used to improve results: first,  suppose that the single contributions are of quite different nature, for example, that they show a different degree of nearsightedness, or that they suggest naturally a perturbation expansion in a different parameter, or around a different zero order. This would allow one to find a better approximation for each contribution alone and hence also to improve the final result in many cases. Even when it is not advantageous, however, the fact that connectors are in principle different for every piece of interest  has to be kept in mind as soon as one is interested, for some reason, in the individual pieces: when the connector $n^c$ is designed to approximate the sum, it could well be that some single pieces are not well described. This would not be a failure of the COT approximation, but of the way it is used. More generally, supposing transferability of the connector is \textit{a priori} an approximation. It is the approximation that we have used for the band structure of helium. That example illustrates the deviations from the exact results, including the appearance of spurious poles. 
	
	\begin{figure}[!t]
		\includegraphics[width=
		0.8\columnwidth]{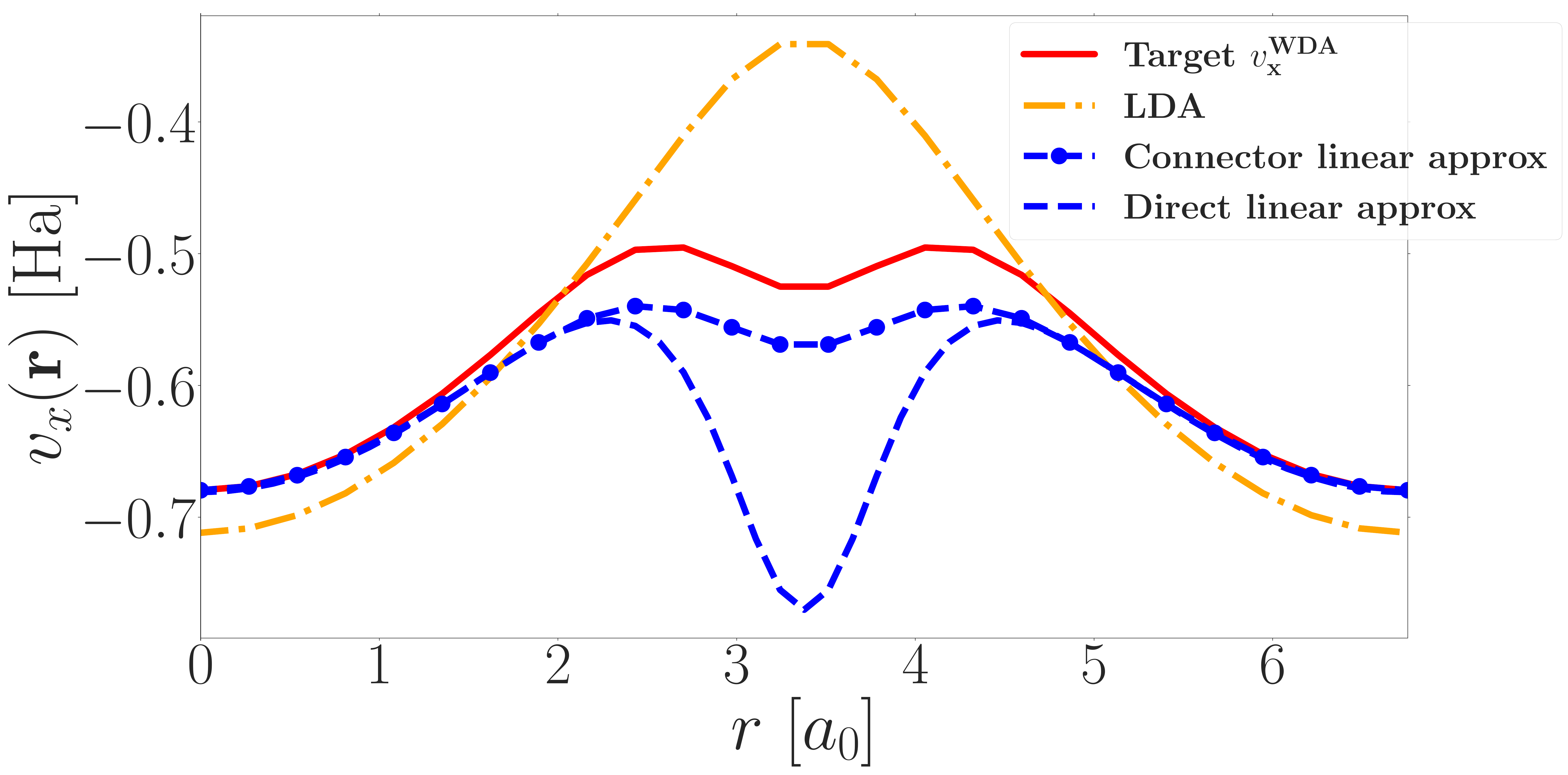}  
		\includegraphics[width=
		0.8\columnwidth]{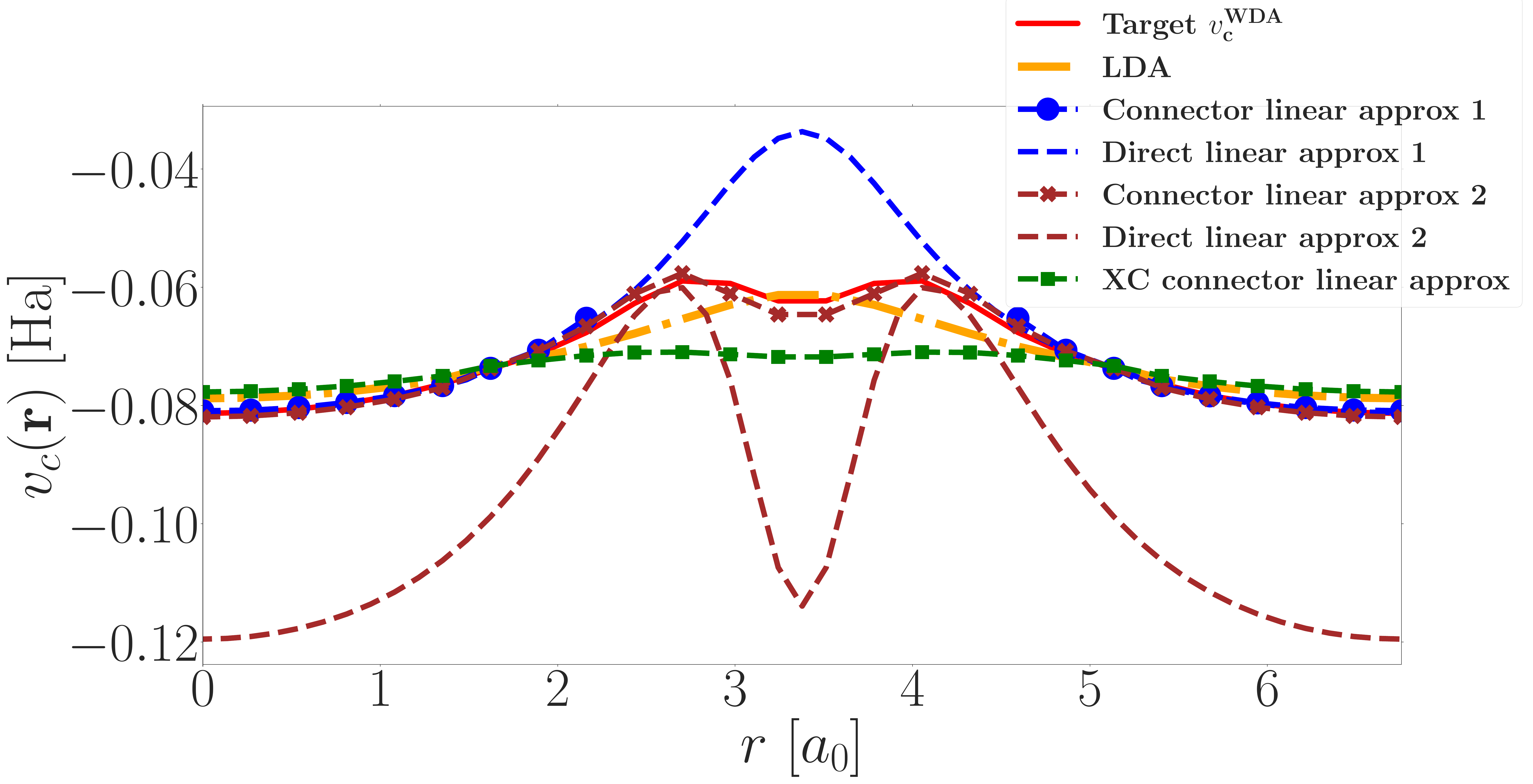} %Gunnarsson_vc_art093.pdf}
		\caption{Comparison between the target WDA  potential (red line) and different approximations: LDA (yellow line), direct linear approximation (dashed blue and brown)  and using the connector (dashed blue and brown, with symbols). Exchange only (upper panel)  and correlation only (bottom panel).  Note that in the correlation potential the connector result is missing  for $r$ in the range $[2.43,4.32]$, since the connector density becomes negative. To solve this problem, it is enough to change the density $n_0$ around which one expands. The dashed brown curve is the result of the direct linear expansion around a density $n_0=n({\bf r})/6$, labeled ``approx 2'', as opposed to ``approx 1'' which stands for $n_0=n({\bf r})$. The brown curve with stars is the corresponding approx 2 connector result. Finally, the green curve with filled squares represents the potential $v_c({\bf r})\approx v_c^h(n^c_{\bf r})$ obtained using the full xc connector, which is not consistent and therefore worsens the result. 
			%\textcolor{blue}{Only if extreeeeemely easy: in the legend, direct before connector :d}
			%\textcolor{red}{Clearly the exchange is the leading contribution in $v_{xc}$...     % 
			% Interesting that the correlation connector doesn't follow the bump in the middle! while the  LDA looks good in that region. The correlation part is under work, because for this picture the correlation was just taken as the difference between $v_{\rm xc}$ and $v_{\rm x}$. If Ayoub manages to make correlation alone work, I will praise the separation to improve results. If he doesn't, I will point out the difficulty to get correlation alone and praise the method to make COT for the sum and for exchange alone, in order to get a - after all not so bad - correlation! BTW, Ayoub, is the direct linear approx for correlation also a difference or is it really directly calculated for the correlation? Could you please show both, just for us now? \textcolor{blue}{For the Linear approximatiom, everthing is linear so making the difference is exactly the same thing as calculating it directly for the correlation  }}      
		}
		\label{fig:Gunn_PZ_exchange}
	\end{figure}
	
	%\clearpage
	
	Since in the case of $v_{\rm xc}$ one is often interested in the exchange and correlation contributions separately, we illustrate this point in Fig. \ref{fig:Gunn_PZ_exchange}, again using the WDA, and now determining the exchange and correlation connectors separately. Here the exchange part is defined by imposing the sum rule on the exchange hole and the correct limit to the HEG, the correlation contribution being the difference to the full WDA. The exchange potential (upper panel) is an order of magnitude larger than the correlation contribution; note the change in scale. The picture for the exchange is therefore very similar to the full xc result. The correlation contribution shows a different behaviour, with the LDA being too shallow.  The correlation connector well reproduces the target $v_c$ for the larger densities. However, for very low densities the linear expansion around the LDA yields negative connector densities, so the connector result (curve with blue filled circles) is absent for $r$ in the range $[2.43,4.32]$. 
	This is not due to a violation of condition \textbf{[A]}, which is indeed fulfilled, but to the fact that the \textit{approximate} result of \eqref{eq:inverse-approx} falls out of the allowed domain, such that \eqref{eq:connector-result} cannot be evaluated. 
	The problem can be overcome in different ways, the simplest one being a change in the approximation. Indeed, the result of the linear approximation depends on the homogeneous density $n_0$  around which one expands. Fig. \ref{fig:Gunn_PZ_exchange} also shows the result of the direct approximation that we obtain by expanding around a much smaller density than before, here $n_0=n({\bf r})/6$ instead of the local density $n({\bf r})$. Contrary to before, now the direct linear expansion of $v_{\rm c}$ has a bump at low densities, but it is exaggerated, and moreover the results at larger densities are very bad. However, now the resulting connector density is always positive, and the connector $v_c$ can be calculated over the whole range. With respect to $n_0=n({\bf r})$, there is very little change at high densities: this demonstrates again the strong error cancelling inherent in the connector approach, making, as a consequence, the result also very stable. In the region of low density, instead,
	we now also obtain an excellent result. It should be noted, however, that in this region the result does depend on $n_0$ (which is what we have used to solve the problem), and further work is needed in order to define a universal prescription for the choice of $n_0$ for very low densities, where $v_{\rm c}$ as function of the density is very steep, and where a bad choice may therefore lead to pathologies such as the negative density above. One promising route, in particular, is the use of exacts constraints.

	Finally, Fig. \ref{fig:Gunn_PZ_exchange} also shows $v_c$ calculated as $v_{\rm c}({\bf r})\approx v_{\rm c}^h(n^c_{{\bf r}})$, i.e., using for the calculation of the correlation part the connector determined for the \textit{full} exchange-correlation potential, with the original choice $n_0=n({\bf r})$. Clearly, 
	the result is much worse. This illustrates our point that there is a connector for each object of interest, which implies also that one should use separate connectors if one is interested in single contributions. This is not a problem of the approach, because it does not add significant workload; one simply has to be aware of it. 
	
	%\begin{figure}[t]
	%  \includegraphics[width=
	% 0.8\columnwidth]{Gunnarsson_xc_in_vx_a093.pdf}  
	%  \includegraphics[width=
	%  0.8\columnwidth]{Gunnarsson_xc_in_vc_a093.pdf} 
	%\caption{    \textcolor{blue}{The magenta line is  the exchange potential (upper) and the correlation potential (bottom) using the full xc connector. The dashed blue line is the result of using the exchange connector for exchange and respectively  the correlation connector for correlation. }
	%}
	%  \label{fig:Gunn_PZ_exchange2}
	%\end{figure}

	The fact that the exact connectors representing single contributions to an observable are most likely different, implies that it is meaningless to discuss the relation between these contributions on the basis of a connector derived for their sum. In particular, when using only one connector one cannot require the relation between the contributions to fulfill exact constraints. 
	Only in some limiting cases is the question well posed. For example, one can easily define the limit of extreme nearsightedness, the same for every contribution, i.e. one can unambiguously define the LDA and verify whether it fulfills the virial theorem\cite{Averill1981,Williams1983,Levy1985} linking the correlation contribution to the total energy and the exchange-correlation energy density and potential, which it does, indeed. \cite{Averill1981,Williams1983,Levy1985}
	
	Another consequence of the fact that every observable has its own connector concerns the relation between energy and potential.
	%More generally, it must be stressed again that the connector is in principle different for every object of interest: there is no such thing as the optimal density of a HEG to simulate a given material, only an optimal density of the HEG to simulate, for a given material, a given observable, maybe in a given point and/or at a given frequency. Here, for example, we have discussed the connector that yields a good approximation to $v_{\rm xc}$. This does not mean that the same connector will also yield a good approximation to $\epsilon_{\rm xc}$, and hence to the total energy.
	Suppose that we have succeeded to find the connector $n^c_{{\bf r},e}$ that correctly describes the energy density, so
	\begin{equation}
		%\label{Exc_con} 
		E_{xc}[n]\equiv \int d {\bf r} n({\bf r}) \epsilon_{xc}(n^c_{{\bf r},e})\,.
		\label{eq:exc-con}
	\end{equation}
	We now calculate the potential, which is
	the functional derivative of  $E_{xc}[n]$ in \eqref{eq:exc-con}, which yields 
	\begin{eqnarray}
		v_{\rm xc}({\bf r})&\equiv&   \frac{\delta E_{xc}[n]}{\delta n({\bf r})} = 
		\int d {\bf r'}  \frac{\delta \left( n({\bf r'}) \epsilon_{xc}(n^c_{{\bf r}',e}) \right) }{\delta n({\bf r})} 
		\nonumber\\
		& =&  \int d  {\bf r'}  \Bigg[ {\bf \delta(r-r')}  \epsilon_{xc}(n^c_{{\bf r}',e}) %\right. \\ &  \left. 
		+n({\bf r'})  \frac{d \epsilon_{xc}(n^c_{{\bf r}',e}) }{d n^c_{{\bf r}',e} } \frac{\delta n^c_{{\bf r}',e}}{\delta n({\bf r})}   \Bigg]\nonumber\\
		& = & \epsilon_{xc}(n^c_{{\bf r},e}) +  \int d {\bf r'}   n({\bf r'})  \frac{d \epsilon_{xc}(n^c_{{\bf r}',e}) }{d n^c_{{\bf r}',e} } \frac{\delta n^c_{{\bf r}',e}}{\delta n({\bf r})}  \,,
		\label{eq:deri-vxc}
	\end{eqnarray}
	while the connector xc potential using the same connector would read 
	\begin{equation}
		\label{con vxc in func of exc}
		v^c_{xc}({\bf r}) =  \epsilon_{xc}(n^c_{{\bf r},e}) +    n^c_{{\bf r},e}  \frac{d \epsilon_{xc}(n^c_{{\bf r},e}) }{d n^c_{{\bf r},e}} \,,
		%\label{eq:conn-pot}
	\end{equation}
	which is a different expression.
	The connector density that correctly represents $v_{\rm xc}^c({\bf r})$, instead, is obtained by inverting the HEG $v_{\rm xc}^h$ applied to \eqref{eq:deri-vxc} and reads
	\begin{eqnarray}
		n^c_{{\bf r}}&=&v_{\rm xc}^{h,-1} \Big(\epsilon_{xc}(n^c_{{\bf r},e}) +  \int d {\bf r'}   n({\bf r'})  \frac{d \epsilon_{xc}(n^c_{{\bf r}',e}) }{d n^c_{{\bf r}',e} } \frac{\delta n^c_{{\bf r}',e}}{\delta n({\bf r})} 
		\Big )\,.
	\end{eqnarray}
	%which may be different from
	%\begin{equation}
	%    n^c_{{\bf r},e} = \epsilon_{\rm xc}^{-1}\Big (\epsilon_{\rm xc}({\bf r};[n])\Big )\,.
	% \end{equation}}
	%The two connector densities fulfill $n^c_{\bf r}\approx n^c_{{\bf r},\epsilon}$ only under special conditions, for example, 
	% if we had $n^c_{{\bf r}',\epsilon}\approx n^c_{{\bf r},\epsilon}$ in the integral.

	% \textcolor{red}{I suggest to skip the rest of the paragraph. DO YOU AGREE?} \textcolor{magenta}{OK} \textcolor{blue}{A: for me ok with both possibilities. the rest is also interesting }
	%Vice versa, one could start from the connector approximation to $v_{\rm xc}$ and perform a functional integration to obtain the energy, for example, as suggested in [Cruz Lam Burke J. Phys. Chem. A 1998, 102, 4911-4917]. Again, there may be questions for which such a consistent picture is mandatory, but in general it will not necessarily yield a better approximation for the observables, than approximating the energy and the potential separately. A good approximation for the potential does not even have to be a functional derivative at all \textcolor{red}{do we have a good reference? there were these people choosing the integration path when vxc is not a functional derivative.....} \textcolor{magenta}{some MGGA potential only? R. van Leeuwen also energy from potential?}.
	
	With these clarifications in mind, we can resume what our illustration is meant to yield and what can instead not  be inferred: choosing an object of interest (here, $v_{\rm xc}$), we have shown that COT is a convenient framework to derive a simple approximate functional that reproduces the non-local dependence on the density beyond the LDA for an inhomogeneous system, using knowledge from the HEG. When evaluated in the framework of the WDA the results are highly promising, and therefore 
	future work will be devoted to applications of the expressions to real materials, without the weighted density approximation, and to the self-consistent calculation of the density. Other observables, such as the xc energy, can also be treated within COT, but will require a different connector that will be developed separately.   
	
	\section{Comparison to other model-based approximations}
	
	COT is potentially very powerful because it relies on the use of a model that can easily be calculated and/or tabulated once and for all. Other approximation schemes exist that might be used in the same way,  and it is interesting to make some comparison.  
	We will in the following concentrate on two main routes, namely,
	perturbation theory, and the use of a variational principle. 
	
	\subsection{Perturbation Theory}
	\label{sec:pt}
	
	Perturbation theory (PT) is an expansion around a zero-order starting point,  which can be considered to be a model. The model result as well as its derivatives could in principle be calculated and tabulated as function of some parameters, such that the results for real systems can be reconstructed efficiently\cite{Faraday-Ayoub}. 
	
	To compare PT and COT we suppose that both are based on the same model. Moreover, besides the choice of a model, COT (which is in principle exact) requires the choice of an approximation, and for a straightforward comparison this will be the same PT. In other words, here we will look more in detail at the direct use of PT as compared to its use within COT.   
	
	If $\mathcal O$ is the observable in the model and $\mathcal O^{PT}$ its finite-order  PT approximation, the connector result (\ref{eq:connector-result}) is 
	\begin{eqnarray}
		O^c_x &=& \mathcal O(\mathcal Q_x^{c{,PT}}) = \mathcal O^{PT}(\mathcal Q_x^{c{,PT}}) + \Delta \mathcal O (\mathcal Q_x^{c{,PT}})\nonumber\\
		&=&O^{PT}(x;Q^R)  + \Delta \mathcal O (\mathcal Q_x^{c{,PT}})\,,
		\label{eq:pt-straight}
	\end{eqnarray}
	where $\Delta \mathcal O $ contains the higher orders neglected by the finite order perturbation theory in the model, and $\mathcal Q_x^{c{,PT}}$ is the perturbative connector. 
	%, so
	%$O^{\rm approx}(x;Q^R)$ could be the first terms of a perturbation expansion of $O$, around some $O(Q_0)$. In order to compare with the connector approach, let us suppose that $Q_0$ lies in the domain of the model. From (\ref{eq:inverse-approx}) and (\ref{eq:connector-result}) it is clear that 
	%the connector approach based on a perturbation expansion will yield a result equal to the one of the perturbation expansion itself when $\mathcal O=\mathcal O^{{\rm approx}}$, i.e., when the perturbation expansion to finite order yields the exact result in the model. Otherwise, the connector approximates missing higher order terms using knowledge taken from the model:
	%with $\mathcal O\equiv \mathcal O^{{\rm approx}} +\Delta \mathcal O $, 
	In other words, with respect to the direct PT result, the connector adds higher orders. The approximation consists in evaluating the correction not at the true $Q^R$, but  for an effective parameter set that lies in the model domain and is expressed as  $\mathcal Q_x^{c{PT}}$. Our approximation strategy defines  $\mathcal Q_x^{c{PT}}$, and the
	improvement with respect to straightforward PT, which completely neglects the higher orders, has been 
	illustrated in  Figs. \ref{fig:negative}, \ref{fig:sf} and \ref{fig:ldalimit} above. It should be stressed again that this gain comes without additional computational cost.
	
	Finite-order PT can be optimized by choosing an appropriate zero-order starting point and by adapting  the points where the derivatives are taken. The COT result $O^c_x = \mathcal O(\mathcal Q_x^{c{,PT}})$ can also be interpreted as an optimized zero-order PT calculation.  This means that all other orders, or their sum, should be small. For the first order, this requirement reads
	\begin{equation}
		\int \left. \frac{dO(x)}{dQ}\right|_{Q=Q_0} (Q^R-Q_x^c) = 0\,,
	\end{equation}
	where the integral symbolizes integration and/or summation over parameters, and where we have allowed for the possibility that the derivative is taken at some optimized parameter set $Q_0$ within the model domain. 
	Solving for the optimized zero order $Q_x^c$ by supposing, as throughout this work, that the model can be represented by one effective parameter $\mathcal{Q}$, one obtains  
	\begin{equation}
		\mathcal{Q}_x^c = \Big (\int \left. \frac{dO(x)}{dQ} \right|_{Q=Q_0} Q^R \Big )/\Big ( \left. \frac{d \mathcal{O}(x)}{d\mathcal{Q}}\right|_{\mathcal{Q}=\mathcal{Q}_0}\Big )\,,
	\end{equation}
	which is right the first order connector. In other words, supplementing PT with COT can be seen, alternatively, as a way to efficiently simulate higher orders, or as a way to optimize the zero-order by making some higher orders disappear. The latter point of view indicates that COT based on PT will work well if making some low order (in our example: the first order) disappear also leads to vanishing or small higher orders. This suggests one possible way to analyze a given problem in order to optimize the COT approximation, for example, through the choice of $\mathcal{Q}_0$. 
	
	Finally, it should again be stressed that a perturbation expansion is only \textit{one possible realization} of a connector approximation, and COT can also be used to improve completely different approximations. The cutoff on the Coulomb interaction in Fig. \ref{fig:sf} is a striking example.
	
	\subsection{Variational search}
	
	Designing a model system such that it represents certain aspects of a real system can be seen as an optimization problem, which allows one, in some cases, to use a variational principle. In particular, when one is directly interested in energies, this can be used straightforwardly, since the model system can be seen as a trial approximation to the real system and tuned such that the energy is minimized. For other observables this is not necessarily true: one may hope that the model configuration that minimizes the energy  also well describes other observables of the system, but there is no guarantee. The connector strategy, instead, applies equally to all observables.
	
	Still, it is interesting to compare the variational and the connector approach for the calculation of an energy. We will use the example of the 1D potential of Sec. \ref{sec:1Dpot} for a simple comparison of principle and to give numerical results.
	
\begin{figure}[!t]
		\includegraphics[width=
		0.8\columnwidth]{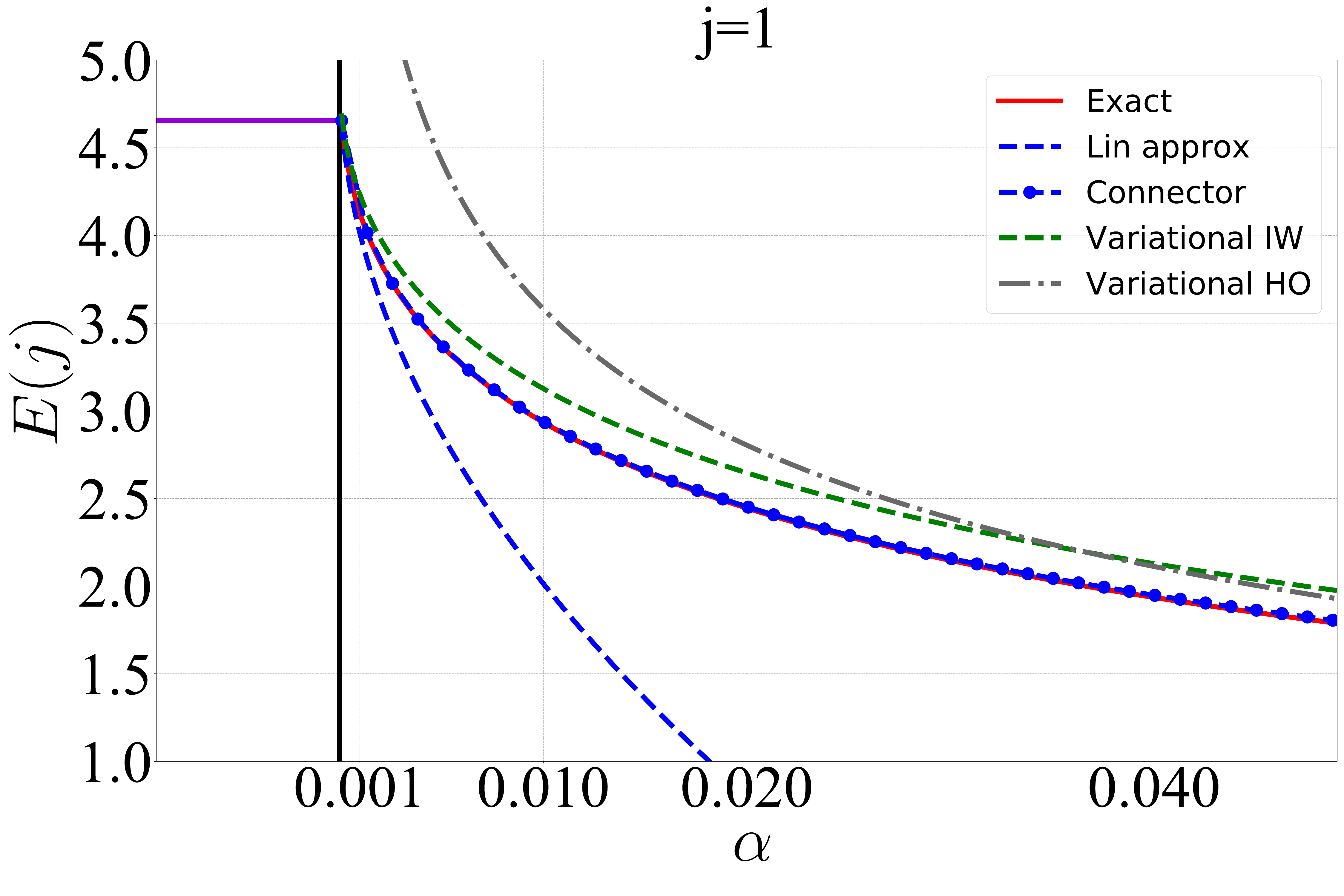}  
		\includegraphics[width=0.8\columnwidth]{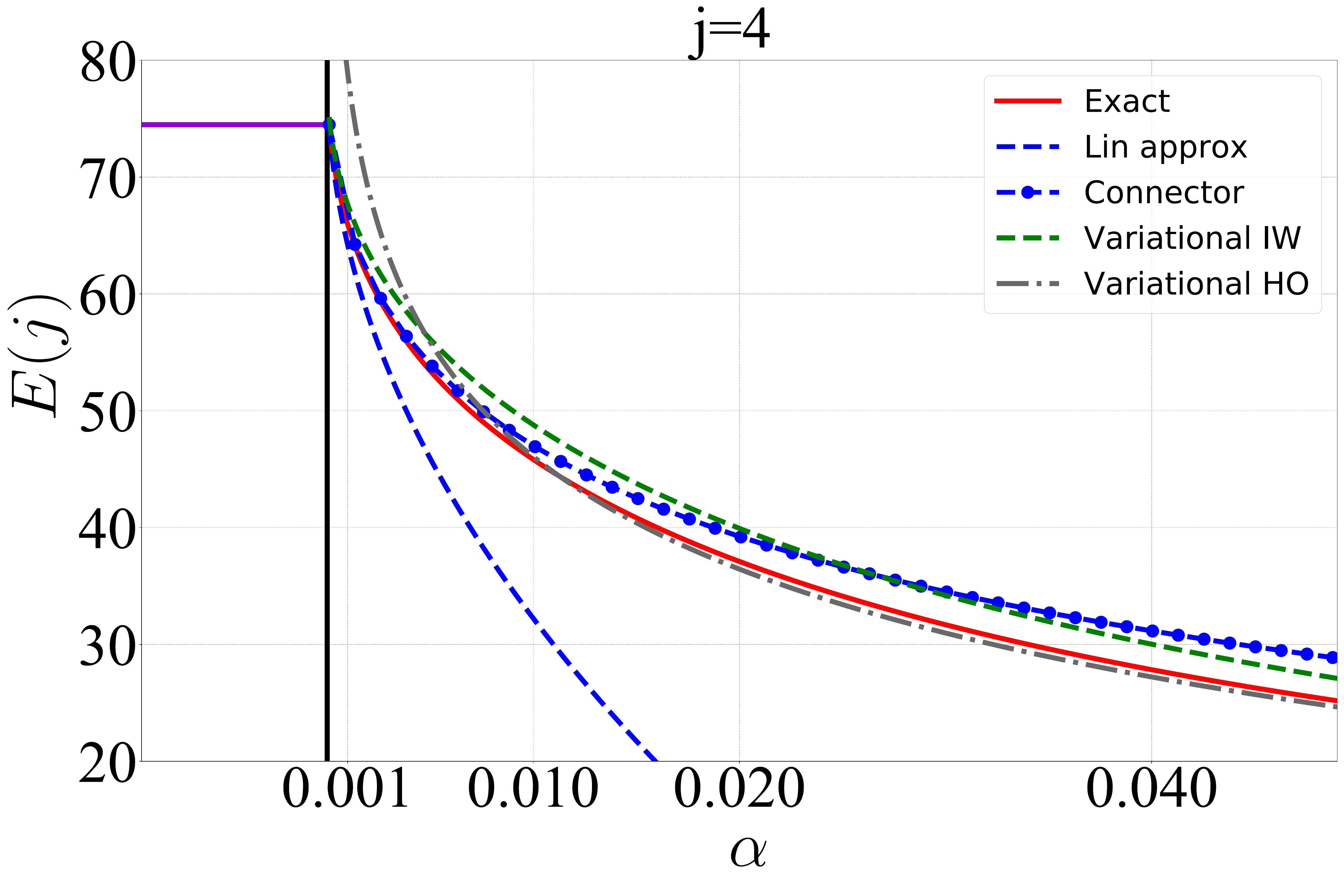}
		\caption{  Energy levels $E(j)$ in the one dimensional system of Fig. \ref{fig:negative} for (upper panel) $j=1$ and (lower panel) $j=4$.   
			%Energy levels $E(j)$, multiplied by $L^2$  , for the 1-dimensional system of Fig. \ref{fig:negative} for (upper panel) $j=1$ and (lower panel) $j=4$. The solid red line is the exact result and the blue dashed line $E^{\rm approx}_j$ of first-order perturbation theory. 
			%    of the equation of motion. 
			The solid red line is the exact result and the blue dashed line $E^{\rm approx}_j$ of first-order perturbation theory. Blue dashed line with symbols: connector results $\mathcal E_j(L^{c,approx}_j)$ corresponding to the first-order perturbation theory. Green dashed, variational result using the wavefunctions of an infinite square well (IW) of width $L^v$, corresponding to the level shown in the left panel. The variation optimizes $L^v$ by minimizing the energy. Grey dot-dashed: variational result using the wavefunctions of an harmonic oscillator (HO) of frequency $\omega_0^v$. The variation optimizes $\omega_0^v$ by minimizing the energy. }
		\label{fig:vari}
	\end{figure}

	The strategy of the variational approach is to minimize the energy of the real system by varying the parameters of the wavefunctions of the model system. For our model system,  the infinite potential well of width $L^v$,
	the wavefunctions are
	%\begin{eqnarray}
	%    \phi_{j=2n}(z)&=&\sqrt{\frac{2}{L^v}}\,\sin(\frac{z}{L^v}j\pi)\,\,\,\,\,\,\,\,\,\,\,\,\,\,\,\,\,\,-\frac{L^v}{2}<z<\frac{L^v}{2}\nonumber\\
	%     \phi_{j=2n+1}(z)&=&\sqrt{\frac{2}{L^v}}\,\cos(\frac{z}{L^v}j\pi)\,\,\,\,\,\,\,\,\,-\frac{L^v}{2}<z<\frac{L^v}{2}\nonumber\\\phi_j(z)&=& 0\,\,\,\,\,\,\,\,\,\,\,\,\,\,\,\,\,\,\,\,\,\,\,\,\,\,\,\,\,\,\,\,\,\,\,\,\,\,\,\,\,\,\,\,\,\,\,\,\,\frac{L^v}{2}<|z|\,,
	%\end{eqnarray}
	\begin{alignat}{3}
		\phi_{j=2n}(z)&=\sqrt{\frac{2}{L^v}}\,\sin(\frac{z}{L^v}j\pi) \,\,\,\,\,\,\,\,\, && -\frac{L^v}{2}  <z &&<\frac{L^v}{2}\nonumber\\
		\phi_{j=2n+1}(z)&=\sqrt{\frac{2}{L^v}}\,\cos(\frac{z}{L^v}j\pi) \,\,\,\,\,\,\,\,\, && -\frac{L^v}{2}  <z && <\frac{L^v}{2}\nonumber\\
		\phi_j(z) &=  0 && \,\,\, \frac{L^v}{2}  <|z| &&\,.
	\end{alignat}
	%since 
	%\begin{equation}
	%\begin{multline}
	%    \int_{-\frac{L^v}{2}}^{\frac{L^v}{2}} dx\,\sin^2\left(\frac{x}{L^v}2n\pi\right)= \\ \int_{-\frac{L^v}{2}}^{\frac{L^v}{2}} dx\,\cos^2\left(\frac{x}{L^v}(2n+1)\pi\right)=\frac{L^v}{2}\,.
	%    \end{multline}
	%\end{equation}
	The kinetic energy in a state with even or odd $j$ is
	\begin{equation}
		E^{\rm kin} = \frac{j^2\pi^2}{2(L^v)^2}\,,
	\end{equation}
	equal to the total energy of the infinite square well with width $L^{v}$. 
	The potential energy is calculated with the potential $V(z;\omega_0,L)$ of the real system with flat region of width $L$, so for states with odd $j$
	\begin{equation}
		E^{\rm pot}=\frac{2}{L^v}\int_{\frac{L}{2}}^{\frac{L^v}{2}} dz\,\cos^2\left(\frac{z}{L^v}j\pi\right)\,	\omega_0^2\left(z-\frac{L}{2}\right)^2\,,
	\end{equation}
	where $L^v>L$,  and where the upper integral boundary is due to the fact that the trial wavefunction vanishes outside the region defined by $L^v$. An equivalent expression holds for even $j$. We now minimize $E^{\rm kin}+E^{\rm pot}$ with respect to $L^v$, for given $j$, $L$ and $\omega_0$. This has to be done numerically. The result is shown in Fig. \ref{fig:vari}, for the lowest level  $j=1$ (upper panel) and for $j=4$ (lower panel). The variational method performs very well in both cases. For $j=1$, the connector outperforms the variational approach. For higher $j$, the connector does better than the variational approach for small $\alpha$ and worse for larger $\alpha$, as shown here for  the case $j=4$ (lower panel), where the crossover is at $\alpha\approx 0.02$. The crossover point decreases with increasing $j$.
	
	Both the crossover and the fact that the variational method tends to win earlier for higher levels can be understood by the fact that we have chosen a first order expansion for the connector approximation: this approximation becomes worse for larger curvature, and eventually the error cancelling in the connector approach, though impressive, is not efficient enough to repair the bad perturbation result.
	Also the trend with increasing $j$ can be understood:
	in our example  
	we have, for all levels, expanded around a box of width $L_0=L$. This is a good choice for lower levels, but higher levels experience an increasingly larger effective width, as suggested by Fig. \ref{fig:negative}. This explains why the direct first order approximation becomes worse for higher levels (note the change in scale between upper and lower panel), and explains the earlier crossover of connector and variational results.
	
	The variational approach adapts automatically to each situation. 
	In the connector approximation based on perturbation theory, the expansion point (here, chosen to be $L_0=L$) introduces some flexibility.  In the present example, one can use for an efficient expansion of higher levels a larger $L_0$, relying on the fact that improving the direct approximation  also improves the connector result. Of course, for the method to be predictive such a further optimization must stem from fundamental principles, e.g., fulfillment of exact constraints, or self-consistency.
	%, for example, by using the classical turning point at the energy $\varepsilon_j^{\rm box}$ instead of the width $L$ at the bottom of the potential. This would improve the direct perturbation approximation, and therefore the connector result, without adding computational cost. 
	
	It is important to note that the computational cost for the variational approach is clearly higher than for the perturbative connector approximation. 
	This suggests that  within a significant parameter range it may be much more efficient to use COT based on a perturbation expansion, than using the variational approach. As our simple example shows, this holds even for the calculation of energies, which are naturally variational, contrary to most other observables.
	
	Finally, we will use our simple 1-dimensional example to illustrate another aspect of COT, linked to the choice of the model system. As pointed out since the beginning, for COT to be efficient the approximation must be equivalent for the real and the model system. Now suppose we had chosen another model system, namely, the harmonic oscillator, which corresponds to $\alpha\to\infty$. This model can be used, for example, in the variational approach, with the results shown in Fig. \ref{fig:vari}: as expected, they are very good for large $\alpha$ and deviate significantly from the exact result for small $\alpha$. Note that now the performance of the variational approach is better for higher levels, since they are less influenced by the flat bottom of the potential. The harmonic oscillator can also be used as starting point for a perturbation series, where we expand the real system around $L=0$ which yields excellent results for large $\alpha$ and becomes very bad for small $\alpha$, much sooner than the variational result\cite{federico}: in the range of $\alpha$ chosen in Fig. \ref{fig:vari}, the results would be too negative to be seen in the figure. One would now like to improve the perturbation result by using COT. To do so, one has to find an equivalent approximation in the model, i.e. the oscillator. The only parameter here is $\omega_0$.  However, the energy levels of the harmonic oscillator are linear in $\omega_0$, which means, the first order expansion in the model would already be exact, contrary to the first order expansion in the real system. Therefore, in this case COT would not benefit from error cancelling, but simply yield the same result as direct perturbation theory, so we would not be able to apply the strategy required by condition \textbf{[C]}. In our example, if one is interested in large $\alpha$ one would therefore have to use another approximation to benefit from COT, e.g., an approximation on the kinetic energy.
	
	Indeed, as a final note it is important to remind again that COT can in principle be combined with any approximation, not necessarily perturbation theory: one may even consider to combine it with the variational approach. 
	Therefore, the important question is not so much how to \textit{compare} other approximations to COT, but rather how to \textit{combine} them with COT in the most efficient way.

	\section{Conclusions}
	
	In conclusion, inspired by a historical strategy to approximate density functionals using the homogeneous electron gas, we propose an in principle exact and very general approach. Its aim is to calculate once and for all, and store, a given observable or other object in a model system with high precision. These results are then used to determine the same object in real systems, \textit{via} a procedure termed connector. The connector is different for every target object, and must be approximated. We suggest a strategy for a systematic connector approach, which, for a given model, makes use of an approximation in a way that leads to strong error cancelling. Indeed, we show that, and why, a given approximation is often much more powerful when used within connector theory than when directly applied to the object of interest.  Of course, the choice of the model and of the approximation still determine the quality of the results and require care and physical insight.  
	The approach opens the way for fast computational methods, which we have illustrated with a quick estimate of energy levels and bands. It can also be used to design functionals. As example, we have derived an exchange-correlation potential for DFT, using as model system the HEG. By using the weighted density approximation as benchmark functional we have demonstrated that our connector functional is able to capture non-local effects in a very efficient way. Beyond the WDA, the connector approximation to the exact functional only requires data from the HEG that are already available in the literature as interpolated expressions. More complex models may require new calculations to be done, with results that can be either interpolated with modern machine learning algorithms or stored thanks to today's storage capacities. Such more flexible models could potentially further improve the results \cite{Faraday-Ayoub} and are envisaged for future applications.
	The present work sets the framework, elucidates the fundamentals and suggests directions for practical application, with a potentially huge impact on computational materials design.
	
	\appendix
	\section{One-dimensional potential} 
	\label{sec:box}
	Here we will give some details concerning the numerically exact solution of the 1-dimensional Schr\"odinger equation with potential \eqref{eq:pot}, and concerning the first-order expansion.
	
	The renormalized Schr\"odinger equation \eqref{eq:sch-scaled} can be solved analytically in each of the regions I, II and III.
	The condition of continuity of the wavefunction between different regions of the potentials yields an equation for an effective wavevector,\cite{federico}
	\begin{equation}
		\tan k=\frac{2kh}{k^2-h^2},
		\label{eq:1}
	\end{equation}
	where $k$ is the standard (region II) and $h$ is a generalized (region I and III) wavevector, defined by:
	\begin{equation*}
		k=\sqrt{2\varepsilon}\qquad
		h=\frac{2}{\sqrt{\alpha}}\frac{\Gamma\left(-\frac{\alpha\varepsilon}{2}+\frac{3}{4}\right)}{\Gamma\left(-\frac{\alpha\varepsilon}{2}+\frac{1}{4}\right)},
	\end{equation*}

\begin{figure}[!t]
	\includegraphics[width=0.8\columnwidth]{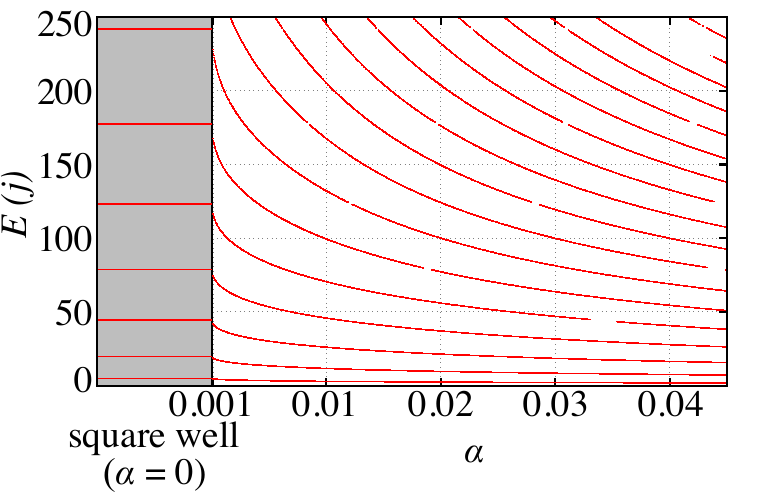}
	\caption{In the main right panel, energy levels $E_j(\alpha)$  as a function of $\alpha$ for $L=1$, from $\alpha^{-1}=10000$ to $\alpha^{-1}=20$. In the left panel, the energies $\mathcal E_j(L)=\varepsilon_j^{\rm box}/L^2$ of the infinite potential well. } 
	\label{fig:exact}
\end{figure}
	
	\noindent {with $\Gamma$ the Gamma function.} The two solutions of Eq. \eqref{eq:1} are $h_{\rm O}=k\tan\frac{k}{2}$ and $h_{\rm E}=-k/\tan\frac{k}{2}$, independent of $\alpha$. %and with $k=\sqrt{2\varepsilon}$.
	Using the definition of $h$ yields two sets of discrete eigenvalues $\{\varepsilon_{\rm O}\}$ and $\{\varepsilon_{\rm E}\}$. Ordering and labelling them with a single index $j\ge1$, one obtains a series where $\{\varepsilon_{\rm O}\}$ correspond to odd and $\{\varepsilon_{\rm E}\}$ to even values of $j$. 
	
	With the definitions for $k$ and $h$ the solution of Eq. \eqref{eq:1} for odd $j$ yields
	\begin{equation}
		\sqrt{2\varepsilon}\tan\sqrt{\frac{\varepsilon}{2}}-\frac{2}{\sqrt{\alpha}}\frac{\Gamma\left(-\frac{\alpha\varepsilon}{2}+\frac{3}{4}\right)}{\Gamma\left(-\frac{\alpha\varepsilon}{2}+\frac{1}{4}\right)}=0\,,
		\label{eq:root}
	\end{equation}
	which we solve numerically for $\varepsilon_j(\alpha)$,
	and analogously for even $j$, 
	using the Ridder method \cite{Ridder} to find the roots. 
	Fig. \ref{fig:exact} shows the exact results; gaps in the red lines indicate {regions} where the root-finder algorithm { was not able to converge}. These gaps were filled in Fig. \ref{fig:negative} by interpolating. This 
	stresses the gain that one may obtain from a more efficient approach, already in this very simple example. From  $\varepsilon_j(\alpha)$  the energy levels are obtained as $E(j;\omega_0,L) = \varepsilon_j(\alpha(\omega_0,L))/L^2$.
	
	In the limit of $\alpha\to0$ one finds $h\to\infty$. Therefore,  Eq. \eqref{eq:root} yields 
	$\sqrt{\varepsilon}\tan\sqrt{\frac{\varepsilon}{2}}\to \infty$. Hence the eigenvalues must fulfill  $\sqrt{\frac{\varepsilon}{2}}=(2n+1)\frac{\pi}{2}$ with $n$ integer, or $\sqrt{2\varepsilon}=j\pi$, $j$ odd integer, and $\varepsilon_j\to \varepsilon_j^{\rm box}\equiv\frac{\pi^2j^2}{2}$. For even $j$ one has analogous relations. 
	To obtain the next order, we make the ansatz $\varepsilon=\varepsilon^{box}_j + c \sqrt{\alpha}$ and expand  $\sqrt{2\varepsilon}\tan\sqrt{\frac{\varepsilon}{2}}$. This yields $c=- \frac{2\Gamma\left(\frac{1}{4}\right)}{\Gamma\left(\frac{3}{4}\right)} \varepsilon_j^{box} $, and finally,
	\begin{equation}
		E^{approx}(j;\omega_0,L)=\frac{1}{L^2}\varepsilon_j^{\rm box}\left[1-\frac{2\Gamma\left(\frac{1}{4}\right)}{\Gamma\left(\frac{3}{4}\right)}\sqrt{\alpha(\omega_0,L)}
		%\frac{1}{\sqrt{L^2\omega_0}}
		\right],
		\label{eq:App2}
	\end{equation}
	which is Eq. \eqref{eq:2}.
	This approximate solution has of course a very small computational cost, compared to the exact one, and presents no numerical difficulties.
	
	\section{Choice of poles in the constrained connector approximation}
	\label{sec:app-poles}
	In order to choose the dominant poles within the scattered connector results, we have to group poles that most probably belong to the same band. To this aim,  the following procedure is applied independently at each ${\bf k}$ (and therefore, does not necessitate any hypothesis about continuation of bands): (i) For each ${\bf K}$, we determine from the spectral function the lowest energy pole, which should in principle be the same for all ${\bf K}$. In practice, one finds a distribution. (ii) We first roughly gather together those poles that lie within ten standard deviations from the average position of the  first pole. These poles appear with different intensities, whereas the true spectral function for non-interacting electrons should only show one, normalized, pole for the first band. Therefore, we sum the intensities of the approximate poles until the normalization is reached; these poles are assigned to the first band. The subsequent poles 
	are assigned to the following band. Within the first group, the pole with maximum intensity is chosen to represent the band $j=1$. (iii) The 
	others 
	are discarded, and the band counter moves on, from $j=1$ to $j=2$. (iv) There may be ${\bf K}$ vectors for which the lowest pole is not visible, because it shows zero intensity.  If the bands are not too dense, the next pole will lie outside of the standard deviation gathering. Also in this case, the counter moves to $j=2$, but those poles will be re-considered. (v) The same procedure is applied to band $j=2$, including as candidate poles the second lowest poles or, for the ${\bf K}$ where (iv) was applied, the poles to be re-considered. The procedure then continues in the same way. Of course, it cannot restore the exact band structure and may fail for large unit cells where bands are very dense, or for materials very far from the HEG where no dominant poles may be identified, and it could certainly be refined. However, it has the advantage of being very simple and fast.

	\section{Weighted Density Approximation}
	
	In the following, we will give more detail concerning our target functional based on the  weighted density approximation (WDA) of the xc hole $n_{\rm xc}$ introduced in \cite{Gunnarsson1977,Alonso1977,Alonso1978,Gunnarsson1979}, with the weight function proposed in \cite{Gunnarsson1980}. The xc energy is given in \eqref{eq:E-WDA}. The two functions $\lambda$ and $C$ are
	\begin{equation}
		\lambda(n) =\left( \frac{-3\Gamma\left(\frac35\right)}{4\Gamma\left(\frac25\right)\varepsilon_{\rm xc}(n)} \right)^5  \text{ and } C(n) = \frac{-3/4\pi}{\Gamma\left(\frac25\right) \left(\lambda(n)  \right) ^{\frac 35} n },  
		\label{eq:constants}
	\end{equation}
	where $\varepsilon_{\rm xc}(n)$ is the exact xc energy per particle of the HEG, whose correlation part we take from the Perdew-Zunger parameterization\cite{Perdew1981}. 
	
	To get the  target xc potential we perform the functional derivative of the xc energy, $\frac{ \delta E_{\rm xc}[n]}{\delta n(\mathbf{r})}$. It yields : 
	\begin{multline} 
		v_{\rm xc}^{\rm WDA}(\mathbf{r},[n]) = %\\ 
		\frac 12 \left[  \int d\mathbf{r'} \frac{ 2 n(\mathbf{r'})}{\mathbf{|r'-r|}} C(\tilde{n}({\bf r'}, {\bf r})) \left(1- e^{\frac{-\lambda(\tilde{n}({\bf r'}, {\bf r}))}{\mathbf{|r'-r|^5}}}\right)  \right.\\
		+ n(\mathbf{r}) \int d\mathbf{r'}    \frac{n(\mathbf{r'})}{\mathbf{|r'-r|}}C'(\tilde{n}({\bf r'}, {\bf r}))  \left(1- e^{\frac{-\lambda(\tilde{n}({\bf r'}, {\bf r}))}{\mathbf{|r'-r|^5}}}\right) \\ \left.
		+ n(\mathbf{r})\int d\mathbf{r'}   \frac{n(\mathbf{r'})}{\mathbf{|r'-r|}^6} C(\tilde{n}(\mathbf{r',r})) 
		\lambda'(\tilde{n}({\bf r',  r}))     e^{\frac{-\lambda(\tilde{n}(\mathbf{r',r}))}{\mathbf{|r'-r|}^5}} \right]\,,
		\label{eq:vxc-wda}
	\end{multline}
	with $\tilde{n}({\bf r}, {\bf r'})= [n({\bf r}) + n({\bf r'})]/2$.\cite{Pablo2000}
	Further functional derivative yields the exchange-correlation kernel, which can be expressed as the sum of six terms, $f_{\rm xc}^{\rm WDA}(|\mathbf{r}-\mathbf{r}'|;n_0) = \sum_{i=1}^6f_i$. Evaluated in the HEG with density $n_0$ these terms read:
	\begin{align}
		f_1 = & \left[  C(n_0) + n_0 C'(n_0) +\frac14 n_0^2 C''(n_0) \right] \frac{1 - e^{\frac{-\lambda(n_0)}{\mathbf{|r-r'|}^5}}}{\mathbf{|r-r'|}}
		\nonumber\\
		f_2 = &  2 \pi \Gamma\left(\frac35\right) \left(\lambda(n_0)\right)^\frac25 \left[ n_0 C'(n_0) + \frac14 n_0^2 C''(n_0)  \right] \delta(\mathbf{r-r'})
		\nonumber\\
		f_3  =&  \left[n_0 C(n_0) \lambda'(n_0) +  \frac12 n_0^2   C'(n_0) \lambda'(n_0) \right. \nonumber \\ 
		& \left. +\frac14 n_0^2  C(n_0)  \lambda''(n_0)\right] 
		\frac{e^{\frac{-\lambda(n_0)}{\mathbf{|r-r'|}^5}}}{\mathbf{|r-r'|}^6}
		\nonumber\\
		f_4  =&  \frac{4 \pi \Gamma \left(\frac35 \right)}{5 \left(\lambda(n_0)\right)^\frac35 }\left[n_0 C(n_0) \lambda'(n_0) +  \frac12 n_0^2   C'(n_0) \lambda'(n_0) \right. \nonumber\\ &  \left. + \frac14 n_0^2  C(n_0)  \lambda''(n_0) \right]    \delta(\mathbf{r-r'}) 
		\nonumber\\
	\end{align}
	\begin{align}
		f_5  =&  -\frac14 n_0^2 C(n_0) \left( \lambda'(n_0) \right)^2 \frac{ e^{\frac{-\lambda(n_0)}{\mathbf{|r-r|}^5}}}{\mathbf{|r-r|}^{11}}
		\nonumber\\
		f_6  =&  - \frac{\pi \Gamma \left(\frac85 \right)}{5 \left(\lambda(n_0)\right)^\frac85} n_0^2 C(n_0) \left( \lambda'(n_0) \right)^2  \delta(\mathbf{r-r'|}) \,.
		\label{eq:fxc-wda}
	\end{align}

	\section*{acknowledgement}
		Stimulating discussions with many members of the Palaiseau Theoretical Spectroscopy Group, and with Ilya Tokatly and Sergio Ciuchi are gratefully acknowledged. This research was supported by a Marie Curie FP7 Integration Grant within the 7th European Union Framework Programme, the European Research Council under the EU FP7 framework program (ERC grant No. 320971), and by the Austrian science Fund FWF under Project No. J 3855-N27. 
	
	%\begin{suppinfo}
	
	%Full details of analytic and numerical calculations for the energy levels of a particle in the one-dimensional box and the band structure of a real material (pdf). 
	%Details on the WDA benchmark for the new exchange-correlation potential (pdf). 
	
	%\end{suppinfo}
	

\end{document}